\documentclass[fleqn,usenatbib]{mnras}

\usepackage{amssymb}
\usepackage{amsmath}
\usepackage{graphicx}
\usepackage{subfig}
\usepackage{xcolor}
\definecolor{forestgreen(traditional)}{rgb}{0.0, 0.27, 0.13}

\usepackage{auto-pst-pdf}

\usepackage{longtable}
\usepackage{url}

\title[MaNGA Photometric and Morphological Catalogs]{SDSS-IV MaNGA PyMorph Photometric and Deep Learning Morphological Catalogs and implications for bulge properties and stellar angular momentum}
\author[Fischer et al.]{\parbox{\textwidth}{J.-L. Fischer$^{1}$\thanks{E-mail: jofis@sas.upenn.edu}, H. Dom\'inguez S\'anchez$^{1}$\thanks{helenado@sas.upenn.edu}, M. Bernardi$^{1}$\thanks{bernardm@sas.upenn.edu} \vspace{0.4cm}}\\
\parbox{\textwidth}{$^{1}$Department of Physics and Astronomy, University of Pennsylvania, Philadelphia, PA 19104, USA\\}}

\begin{document}

\maketitle
\label{firstpage}

\begin{abstract}
We describe the SDSS-IV MaNGA PyMorph Photometric (MPP-VAC) and MaNGA Deep Learning Morphology (MDLM-VAC) Value Added Catalogs. The MPP-VAC provides photometric parameters from S{\'e}rsic and S{\'e}rsic+Exponential fits to the 2D surface brightness profiles of the MaNGA DR15 galaxy sample. Compared to previous PyMorph analyses of SDSS imaging, our analysis of the MaNGA DR15 incorporates three improvements: the most recent SDSS images; modified criteria for determining bulge-to-disk decompositions; and the fits in MPP-VAC have been eye-balled, and re-fit if necessary, for additional reliability. A companion catalog, the MDLM-VAC, provides Deep Learning-based morphological classifications for the same galaxies. The MDLM-VAC includes a number of morphological properties (e.g., a TType, and a finer separation between elliptical and S0 galaxies). Combining the MPP- and MDLM-VACs  allows to show that the MDLM morphological classifications are more reliable than previous work. It also shows that single-S{\'e}rsic fits to late- and early-type galaxies are likely to return S{\'e}rsic indices of $n \le 2$ and $\ge 4$, respectively, and this correlation between $n$ and morphology extends to the bulge component as well.  While the former is well-known, the latter contradicts some recent work suggesting little correlation between $n$-bulge and morphology.  Combining both VACs with MaNGA's spatially resolved spectroscopy allows us to study how the stellar angular momentum depends on morphological type. We find correlations between stellar kinematics, photometric properties, and morphological type even though the spectroscopic data played no role in the construction of the MPP- and MDLM-VACs. 

\end{abstract}

\begin{keywords}
 galaxies: fundamental parameters -- galaxies: photometry -- galaxies: structure
\end{keywords}

\section{Introduction}
The Mapping Nearby Galaxies at Apache Point Observatory (MaNGA; \citealt{Bundy2015}) Survey is a component of the Sloan Digital Sky Survey IV (\citealt{Blanton2017}; hereafter SDSS~IV). MaNGA uses integral field units (IFUs) to map the spectra across $\sim$10000 nearby ($z \sim$ 0.03) galaxies \footnote{At the time of writing, the MaNGA survey is not yet complete: only $\sim$4,700 of the expected $\sim$10,000 galaxies have been observed. The results in this paper refer to the current subset of $\sim 4,700$ objects.}. The IFU technology allows the MaNGA survey to obtain detailed kinematic and chemical composition maps of each galaxy (e.g. \citealt{Gunn06, Drory2015, Law2015, Law2016, Smee2013, Yan1, Yan2, Greene2017, Graham2018}). 
It is interesting to correlate this spatially resolved spectroscopic information with photometrically derived structural parameters of the galaxy. 

\cite{Wake2017} describe how the galaxies were selected from the SDSS footprint for observation. For reasons discussed in \cite{F2017}, we do not use the SDSS pipeline photometry.  However, substantially improved photometry is available through the NASA-Sloan Atlas catalog ({\tt nsatlas.org}; hereafter NSA). This relies heavily on a more careful treatment of object detection, deblending, and the background sky level (see \citealt{Blanton2011} for details). While the NSA photometry provides Petrosian and S{\'e}rsic-based estimates of galaxy magnitudes, sizes, and ellipticities, it does not provide two-component fits. `Bulge-disk' decompositions would be a valuable complement to MaNGA spectroscopy, which provides 2D-maps of rotation and velocity dispersion in galaxies.

In fact, bulge-disk decompositions of about $85 \%$ of the MaNGA galaxies are available in the published catalogs of Simard et al. (\citeyear{Simard2011}; hereafter S11) and Meert et al. (\citeyear{Meert2015}; hereafter M15, catalog referenced as DR7). However, both these analyses were based on SDSS DR7 photometry, which has since been substantially revised. Problems with the estimate of the background sky level are known to have affected the S11 analysis, whereas the results of M15 are less biased compared to, e.g., NSA \citep{F2017}. As the main purpose of our MaNGA PyMorph Photometric Value Added Catalog (hereafter MPP-VAC) is to provide an accurate analysis of the images of MaNGA galaxies which includes the results of two-component fits, and since we would have to analyze the remaining 15\% of the MaNGA galaxies anyway, we thought it prudent to simply re-analyze all the MaNGA galaxies that are currently available. We also provide the SDSS-IV MaNGA Deep Learning Morphology Value Added catalog (hereafter MDLM-VAC) which includes Deep Learning-based morphological classifications (the methodology is described in detail by \citealt{DS2018}) for the same galaxies. The present note describes the main properties of the catalogs and illustrates some of the scientific analysis which they enable.

Section~\ref{sec:pymorph} describes the algorithm we use to determine the photometric parameters listed in the MPP-VAC. The catalog itself is described in Section~\ref{sec:cat}. Section~\ref{sec:lit} compares our photometric parameters with those from previous studies. Section~\ref{sec:morphcat} describes our morphological catalog, MDLM-VAC, and classification. Section~\ref{sec:photmorph} combines our MPP-VAC and MDLM-VAC to show how the photometric parameters correlate with morphology. Section~\ref{sec:spin} combines the MPP-VAC photometry and MDLM-VAC morphologies with MaNGA spectroscopy to study how the angular momentum of galaxies depends on morphological type. A final section summarizes.

\section{MaNGA PyMorph Photometric Value Added Catalog (MPP-VAC)} 

The MPP-VAC is one of the value added catalogs of the SDSS-DR15 release\footnote{www.sdss.org/dr15/data\textunderscore access/value-added-catalogs/} and is available online\footnote{www.sdss.org/dr15/data\textunderscore access/value-added-catalogs/manga-pymorph-dr15-photometric-catalog}. 

\subsection{PyMorph photometry}\label{sec:pymorph}
In what follows, we describe how photometric parameters such as luminosity, half-light radius, a measure of the steepness or central concentration of the profile, etc., were determined by fitting two different models to the surface brightness profiles of MaNGA galaxies: a single S{\'e}rsic profile (\citealt{Sersic1963}, hereafter Ser) and a profile that is the sum of two S{\'e}rsic components (hereafter SerExp). For the SerExp profile, one of the components is required to have $n=1$. It is conventional to refer to the $n=1$ component as the `disk', and the other as the `bulge'. However, later in Section~\ref{sec:flip}, we discuss how this is not always the case.

\subsubsection{Fitting algorithm}
We use a fitting algorithm called PyMorph (\citealt{Vikram2010, Meert2013, Meert2015, Meert2016, Bernardi2014}), a Python based code that uses Source Extractor (SExtractor; \citealt{BA1996}) and GALFIT \citep{Peng2002} to estimate the structural parameters of galaxies. For a galaxy or galaxies in one frame, the image, weight image, and PSF of the image are fed to PyMorph. PyMorph uses SExtractor to define a masked image which is passed to GALFIT which then fits a 2D model to the image. For S{\'e}rsic fits, or for the bulge component of SerExp fits, $n$ cannot exceed 8. When fitting to a two-component SerExp model, there is no requirement that the bulge component be more compact and dominate the light in the inner regions. I.e., it is possible that the algorithm returns a `bulge' that is larger than the `disk'. Since this is not thought to be physically reasonable, we discuss such cases further in Section~\ref{sec:flip}. 


\begin{figure}
  \centering
  \includegraphics[width = 1.0\hsize]{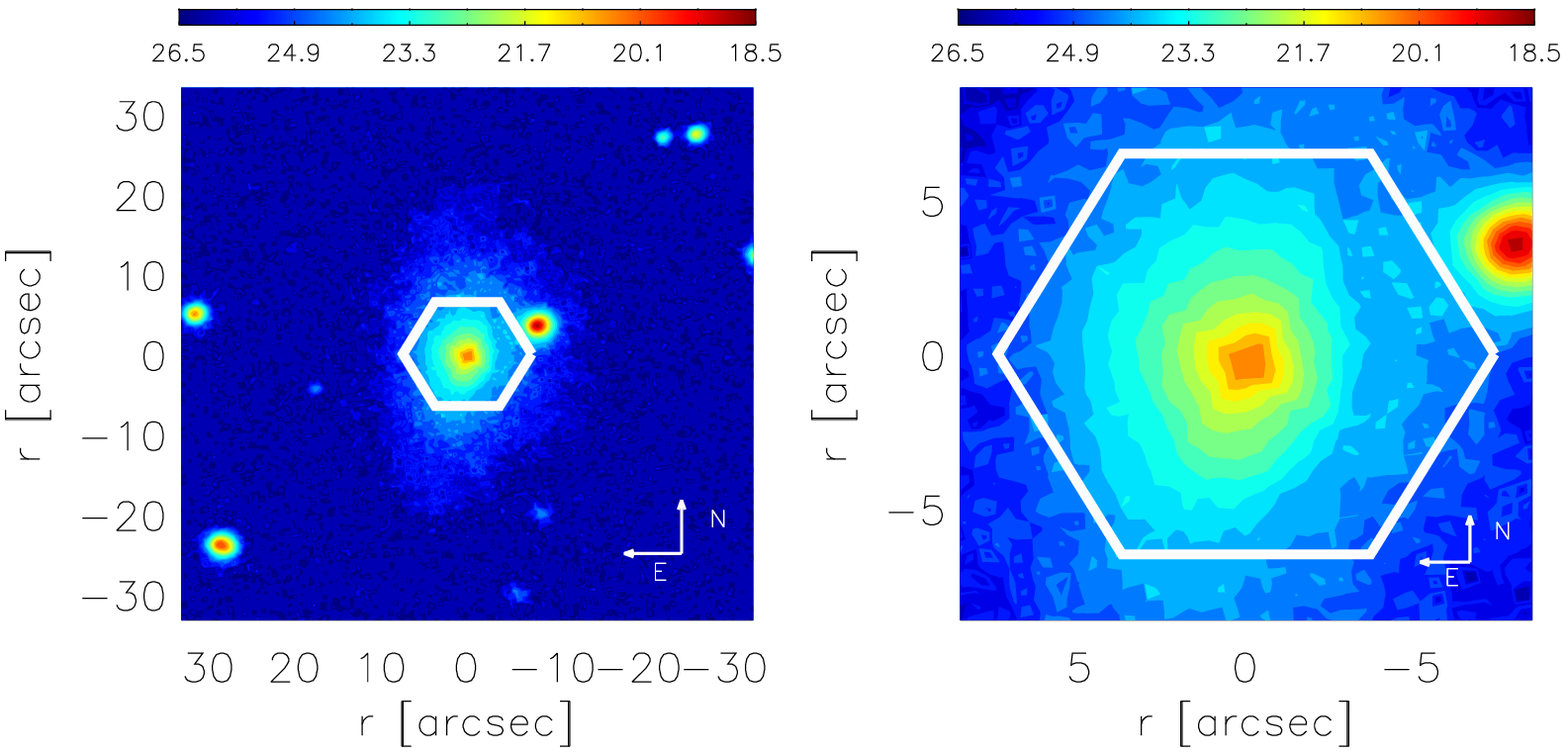}
   \includegraphics[width = 1.0\hsize]{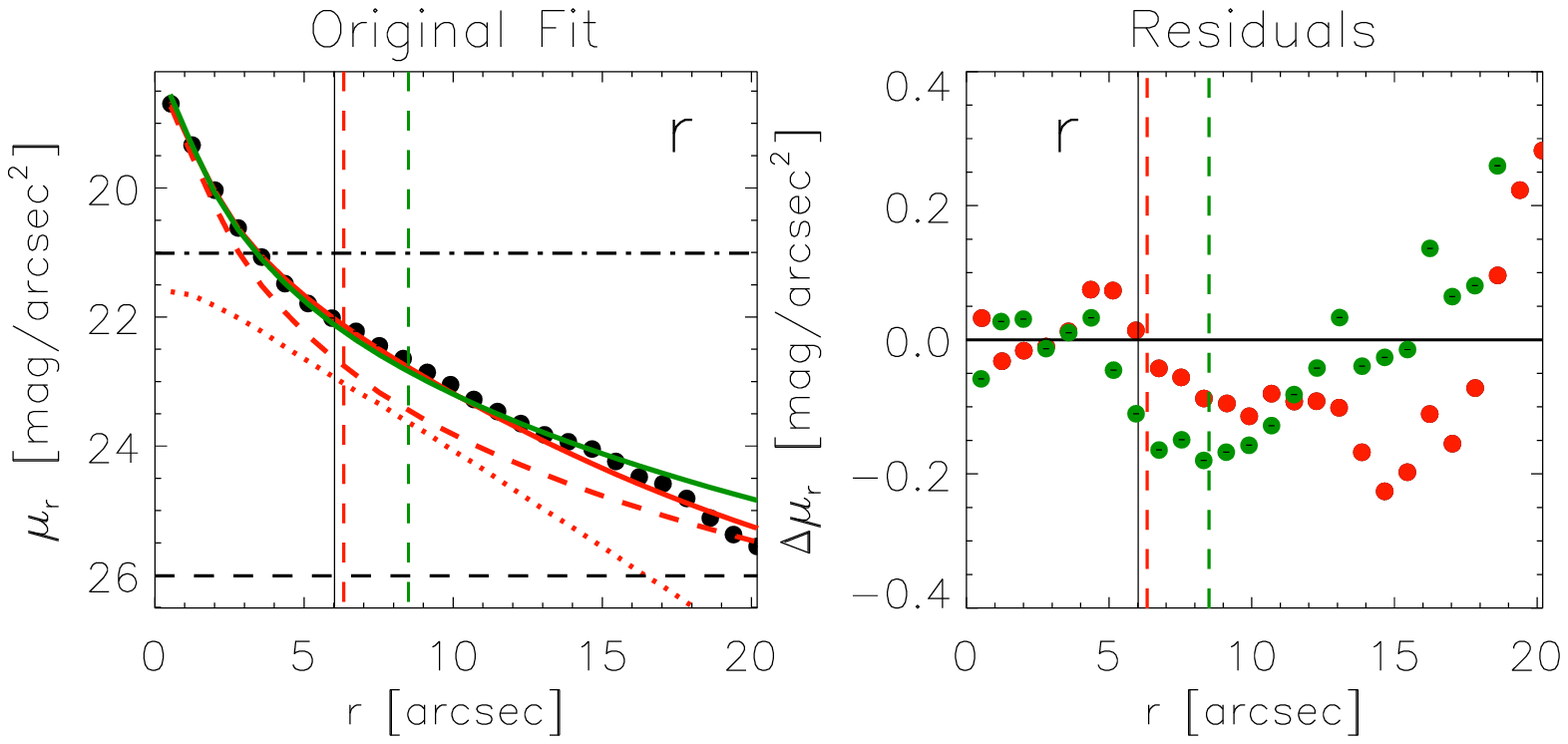}
   \includegraphics[width = 1.0\hsize]{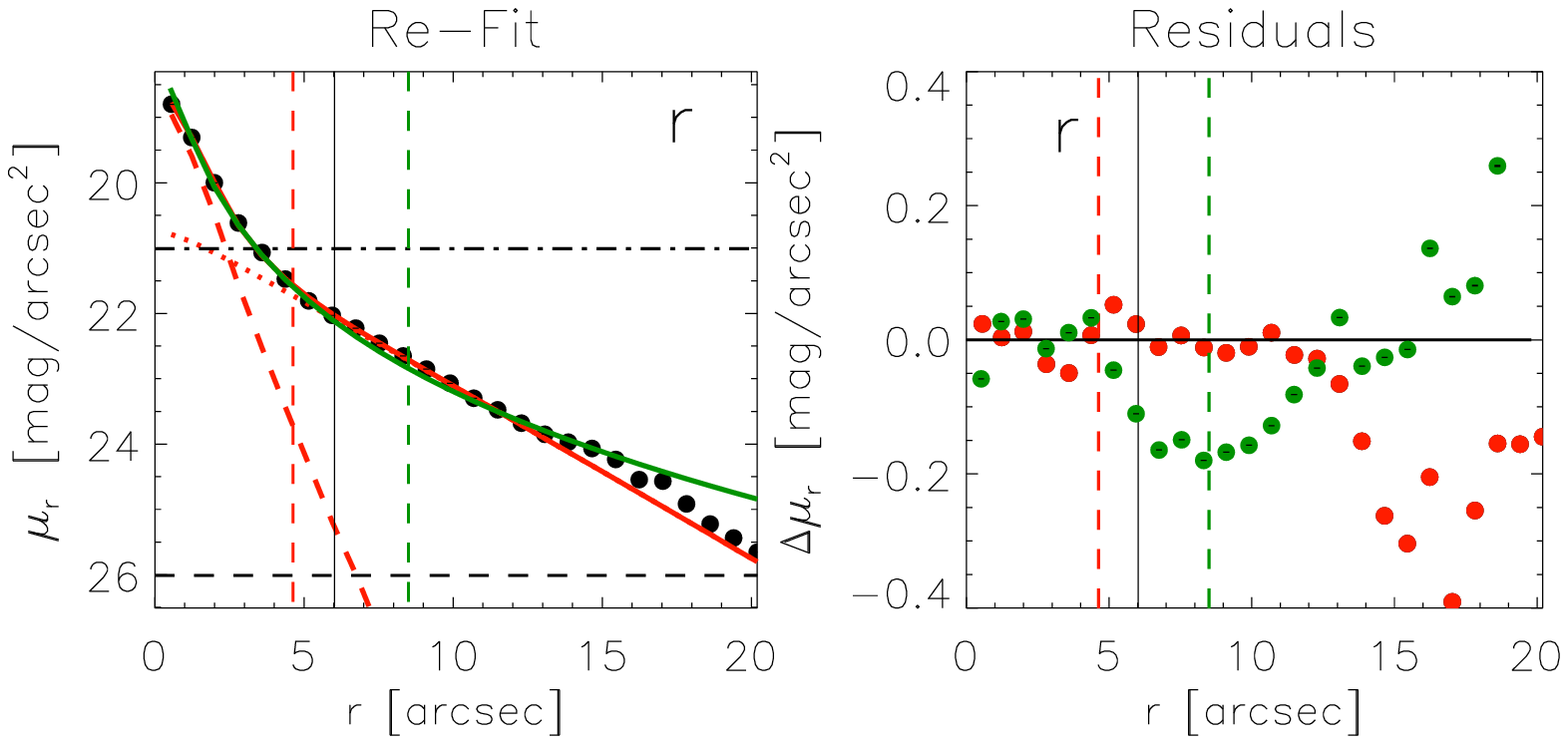}
   \caption{Top left and right: Cutouts of the galaxy image, zoomed in on the right to highlight the area covered by MaNGA IFUs (shown as a white hexagon). The color scale of these images are representative of the surface brightness [mag/arcsec$^2$]. Middle panels are for the original (seeing-convolved) PyMorph fit. Left: Black symbols in the left panel show the 1-d surface brightness profile; solid green line shows the single-component S{\'e}rsic fit; solid red line shows the two-component SerExp fit, which is the sum of an $n=1$ (red dotted) and a bulge component with $n$ as a free parameter (red dashed).  Vertical dashed lines show the associated half-light radii which include half of the total luminosity; vertical solid black line shows the scale covered by MaNGA IFUs. Horizontal lines show the sky level (dash-dotted) and 1\% of sky (dashed). The bulge component has $n\sim 7$, a larger half-light radius than the disk, and dominates on all scales. Right: Residuals from the fits (fits$-$data), SerExp (red) and S{\'e}rsic (green). Bottom panels show the result of refitting after requiring the bulge to have $n<3$ (so only the red SerExp fit curves have changed). The SerExp residuals are substantially smaller, at least within two half-light radii, and the disk now dominates at large radii. 
   } 
\label{fig:refitIII}
\end{figure}

\begin{figure}
  \centering
  \includegraphics[width = 1.0\hsize]{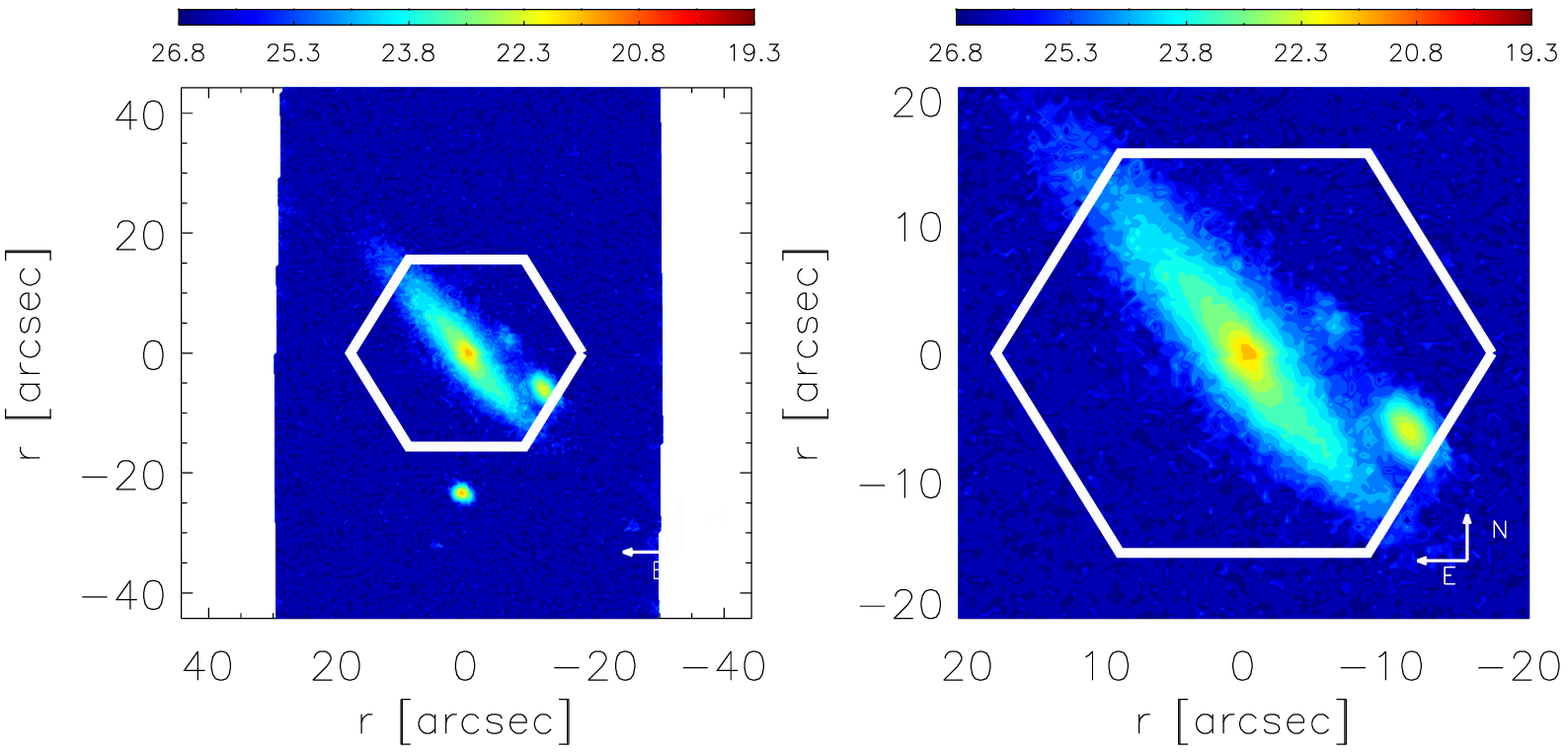}
    \includegraphics[width = 1.0\hsize]{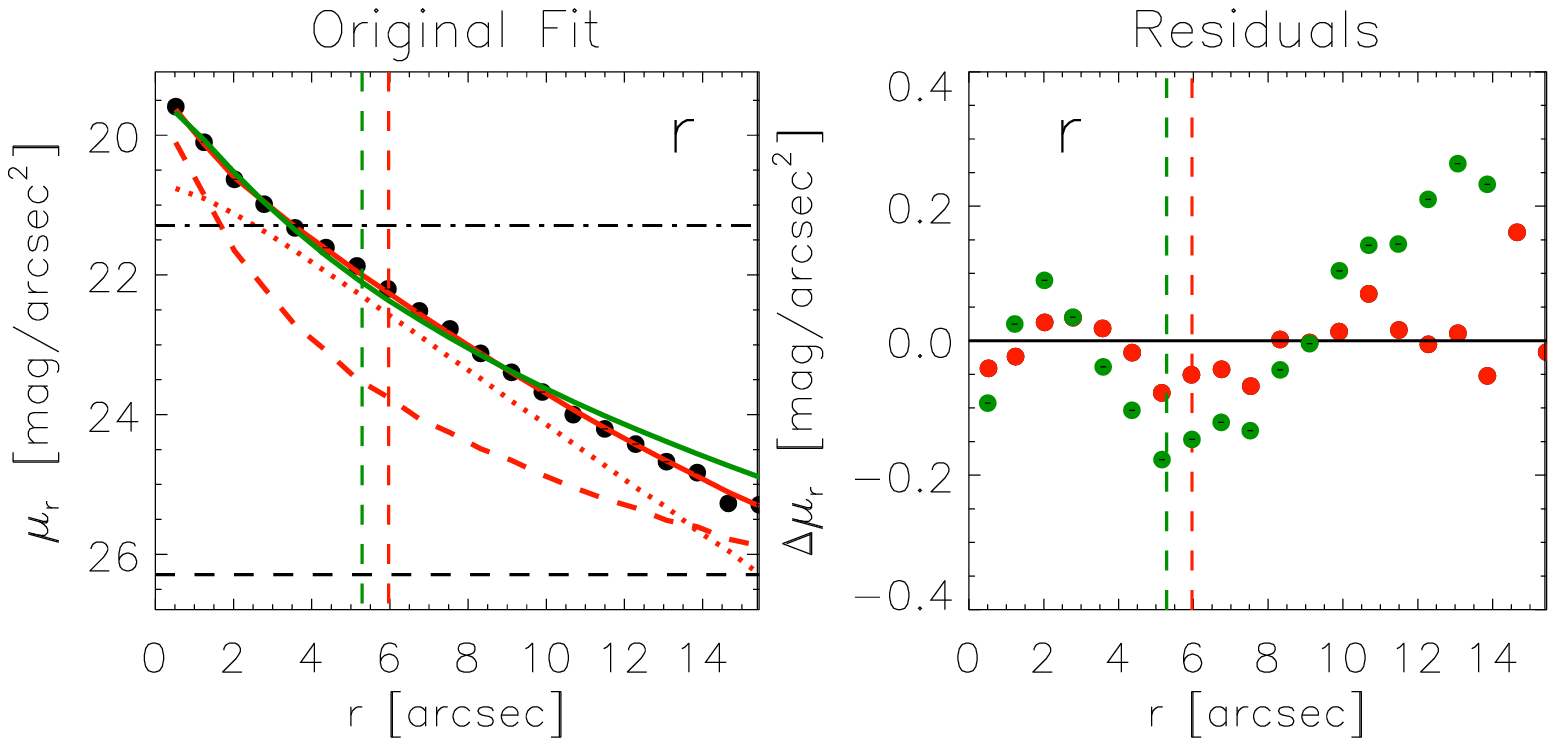}
    \includegraphics[width = 1.0\hsize]{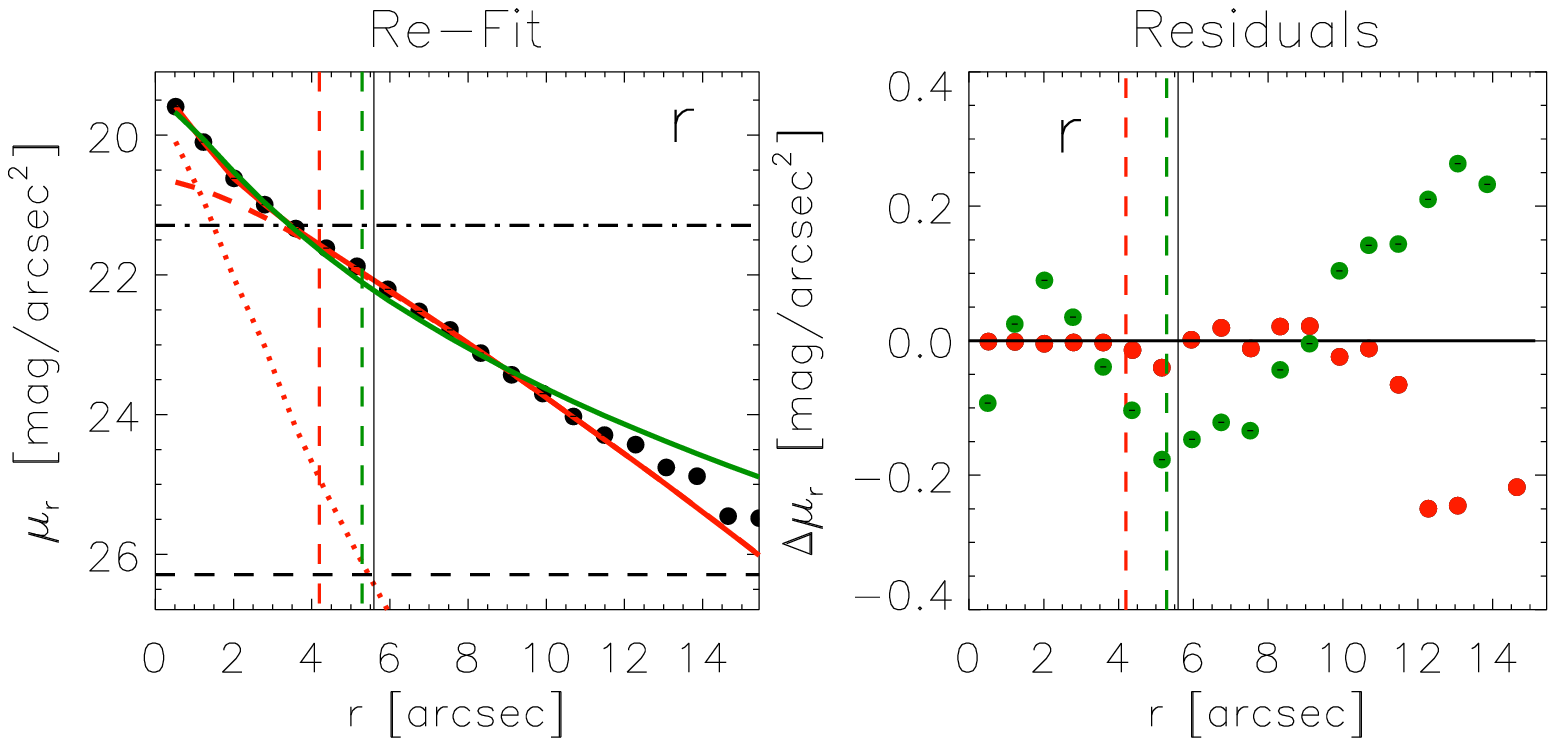}
  \caption{Same as Figure~\ref{fig:refitIII}, but for an object where the original fit returned a bulge with $n>7$ which dominates the light at both small and large radii: the disk only dominates on intermediate scales. Requiring $n<2$ returns a more sensible fit, except that it is `flipped': the $n=1$ component dominates on small scales (a disk would dominate on large scales). In MPP-VAC, this galaxy's parameters are both `re-fit' and `flipped'.}
\label{fig:refitII}
\end{figure}

PyMorph photometric parameters of the galaxies in the SDSS DR7 release \citep{Aba2009} are available online from the {\tt UPenn SDSS PhotDec Catalog} \citep{Meert2015, Meert2016}. As a result, in principle, PyMorph parameters for about 85\% of the MaNGA galaxies are already available. In practice however, these were based on DR7 imaging, which underwent a substantial revision in DR9 and subsequent data releases. Although \cite{F2017} have shown that PyMorph is largely immune to this change, the MaNGA sample is sufficiently small that we thought it prudent to simply rerun PyMorph on the DR15 imaging. 

The small sample size made it possible to also perform a visual inspection of all the objects in the DR15 release of MaNGA. On the basis of this we decided a further re-fit might be justified for some objects. This happens most frequently for the SerExp fit and also most frequently for late-type galaxies in which the bulge component has $n\sim 8$ (but large $n_{\rm bulge}$ is not the only reason). This refitting is described in the following section. 

\subsubsection{Re-fitting}\label{sec:refit}
Some SerExp fits have $n\sim 8$ for the bulge component. Often this is driven by a slight surface brightness excess in the inner most pixel(s), but the resulting bulge component has a long tail to large radii, where it may even dominate over the disk ($n=1$) component. In such cases, we re-run PyMorph, restricting the bulge component to $n\le n_{\rm lim}$ with $n_{\rm lim}=3$ (recall that the default is $n_{\rm lim}= 8$). If the problem persists, then we reduce $n_{\rm lim}$ further to $2$, and finally to $n_{\rm lim}=1$ if necessary.  In effect, this forces the bulge component to dominate in the inner regions only. If the new fit (with smaller $n$) is acceptable (in a $\chi^2$-sense, which we quantify for a few cases below) we keep it and discard the original. While the reduction in $n$ is dramatic (of course), it sometimes (but not always) comes with a similarly dramatic change to $R_{e,{\rm bulge}}$, although the total light and B/T ratio are not strongly affected. Reducing the allowed range in $n$ also has the effect of reducing the effects of degeneracies, thus systematically reducing our error estimates on fitted parameters. Thus, for refitted objects, the uncertainties we report are typically smaller (by $\lesssim 0.1$ mag for the luminosities and $\lesssim 20\%$ for the radii) compared to the error estimates reported for the original fit.  

\begin{figure}
  \centering
    \includegraphics[width = 1.0\hsize]{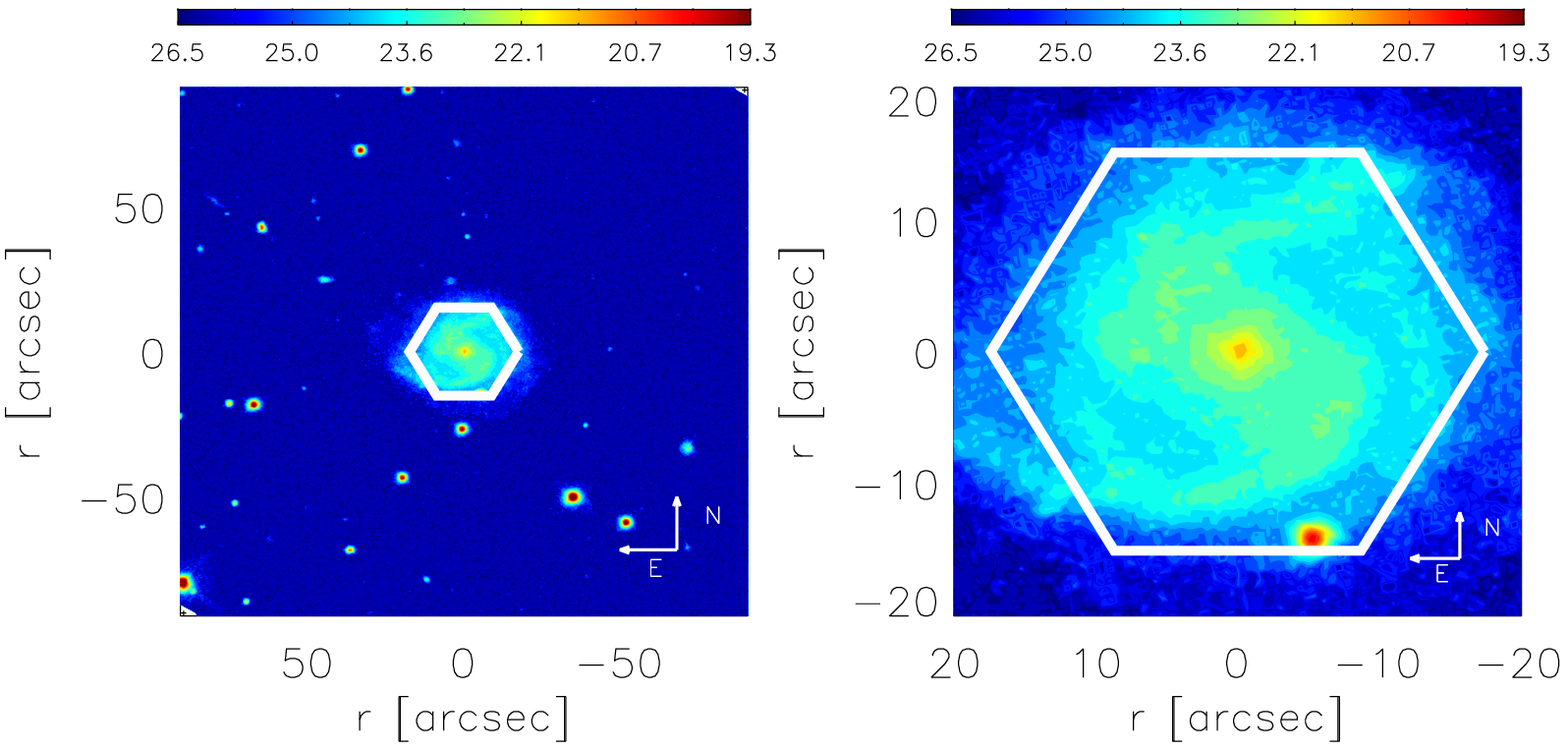}
  \includegraphics[width = 1.0\hsize]{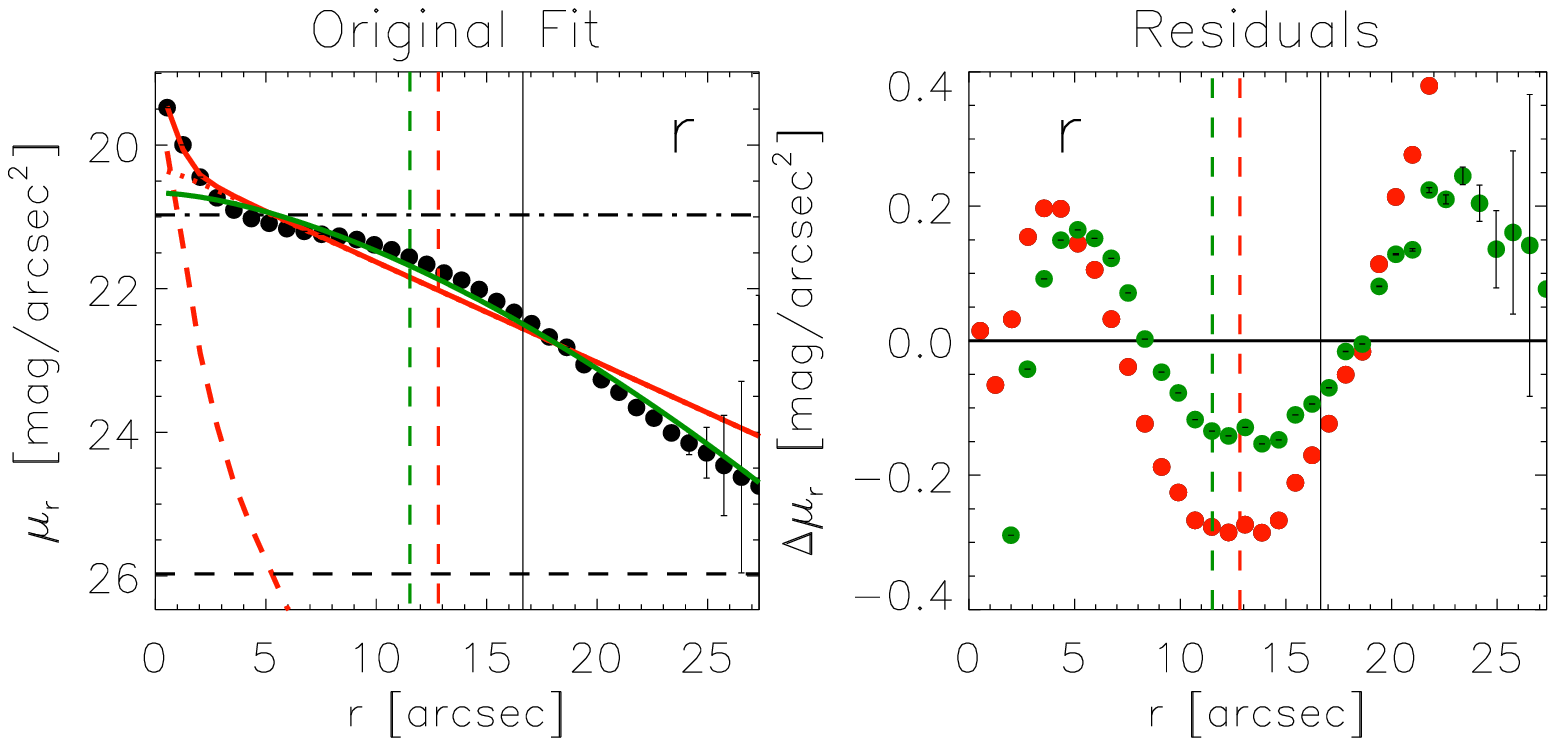}
  \includegraphics[width = 1.0\hsize]{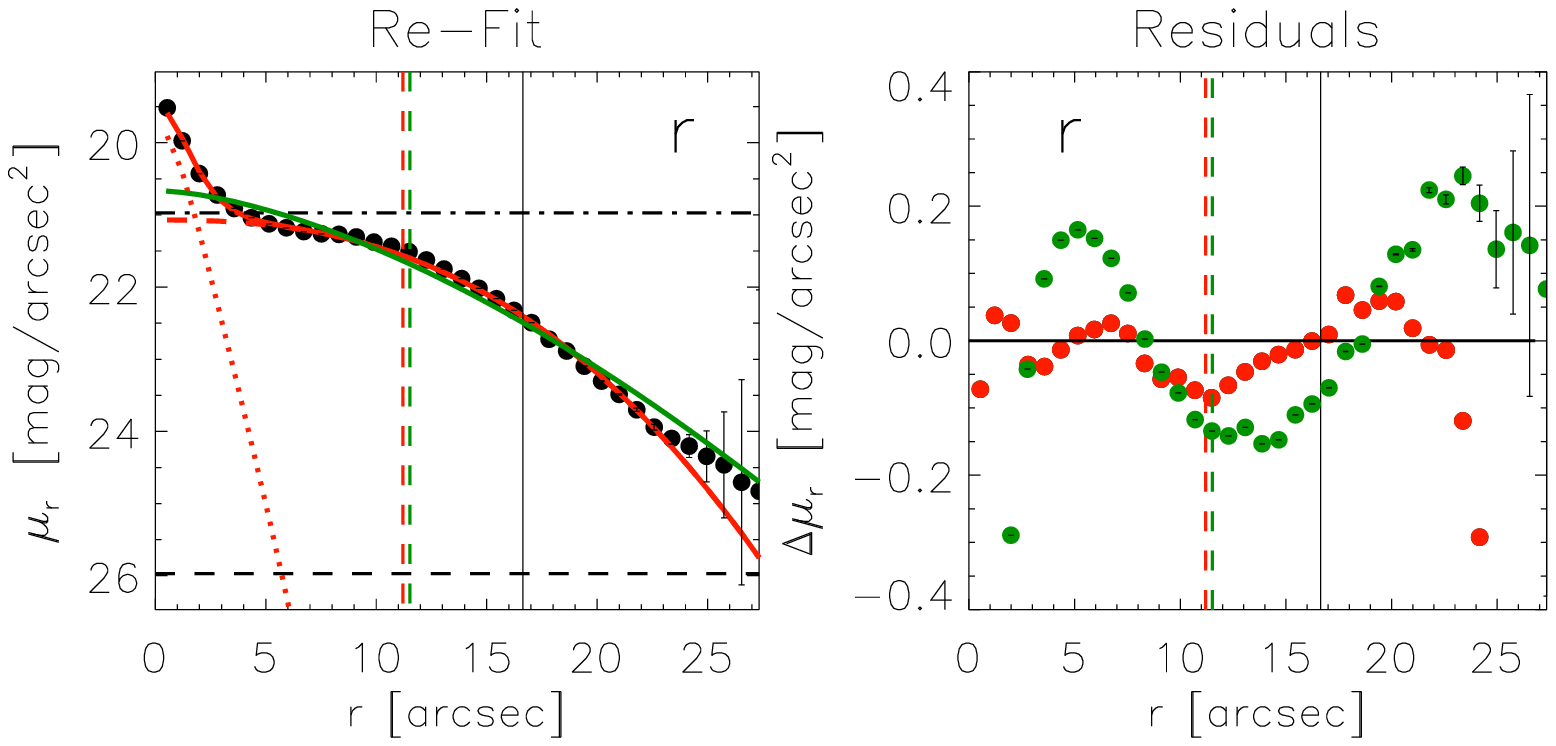}
  \caption{Same as previous figure, but now for an object where the original SerExp fit was simply bad, mainly due to the downturn at large $r$. Requiring the bulge component to have $n<1$ returns a better fit -- one that fits the downturn well -- but again, with `flipped' components, as the $n=1$ component dominates on small scales.}
\label{fig:refitI}
\end{figure}


The following figures illustrate typical examples for when refitting is required. 
Figure~\ref{fig:refitIII} shows a case where the original fit (middle panels) has the bulge (dashed red) dominating the light on all scales, because the SerExp fit returned $n\sim 7$ for the bulge component, and a correspondingly large half-light radius. Requiring the bulge to have $n<3$ and re-fitting (so only the red SerExp fit curves have changed) returns a more compact bulge, so the disk (dotted red) dominates at large radii. Of course, in this case, B/T is also reduced. In this case, $\chi^2_{\rm dof}$ increases from its original value of 1.073 to 1.079; evidently, the $\chi^2$ surface is rather flat.

Figure~\ref{fig:refitII} shows a case in which the original fit had the $n=1$ component (dotted) dominating only on intermediate scales. Requiring the bulge component to have $n < 2$ returns what is essentially two exponential profiles. As a result, the total half-light radius is smaller (reduced from 6'' to 4''). In this case, $\chi^2_{\rm dof}$ increases from 1.1090 to 1.1093.  Notice, however, that the re-fit has the $n=1$ component dominating the inner regions; this is reversed, or `flipped' compared to the usual expectation that the disk dominates the outer parts. We discuss how we report such `flipped' objects in the next subsection.

Figure~\ref{fig:refitI} shows a case in which the original SerExp fit was much worse than the Ser fit. This is mainly because of the obvious downwards curvature at large $r$. Forcing $n<1$ and re-fitting results in a better fit ($\chi^2_{\rm dof}$ decreases from 1.118 to 1.076), although this too is a `flipped galaxy': the $n=1$ component (nominally the disk component) dominates in the inner regions. 

We re-fit $\sim 5$\% and $10\%$ of the MPP-VAC galaxies in the SDSS $r$-band for the Ser- and SerExp-fit, respectively. We find similar fractions for the $g$ and $i$-bands. 

\subsubsection{`Flipped' galaxies}\label{sec:flip}
Some SerExp fits have the $n=1$ component (nominally the `disk') dominating the light in the inner regions, with the $n\ne 1$ component (nominally the `bulge') dominating outside (e.g. Figs.~\ref{fig:refitII} and~\ref{fig:refitI}). 
Cases such as Figure~\ref{fig:refitI}, where the light profile curves sharply downwards at large radii, are typical. For such objects, the best-fit $n$ is always smaller than unity. In such cases, the components of the bulge and disk are `flipped' before being added to the catalog. I.e., after flipping, the inner `bulge' corresponds to the component with $n=1$, and the `disk' component always has $n\le 1$.
In the MPP-VAC $\sim 13$\% of the galaxies are `flipped' in the r-band (similarly for the other bands).

It is conventional to report the half-light radius $R_e$ for the bulge component (usually a S{\'e}rsic profile with $n>1$), but the scale-length $R_d$ for the `disk' ($n=1$) component. Since we sometimes flip the two components, this has the potential for confusion. This is why the radii we report are {\em always} the half-light radius. For $n=1$, the `disk' scale length $R_d$ is related to the half-light scale we report by $R_d = R_e/1.678$ (see equations~4 and 5 in \citealt{Meert2015}).



\begin{figure}
 \centering
 	 \includegraphics[width = 1.0\hsize]{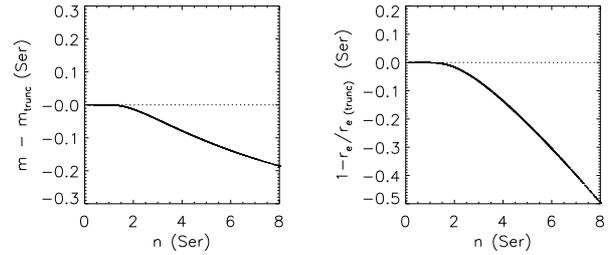}
	  \vspace{-0.5cm}
\caption{Truncation in an ellipse having semi-major axis length $7a_{e(\infty)}$ and axis ratio $b/a$ reduces the total light (left) and size (right) by an amount which depends on S{\'e}rsic index $n$.}
 \label{trunceff}
\end{figure}

\begin{figure}
 \centering
  \includegraphics[width = 1.0\hsize]{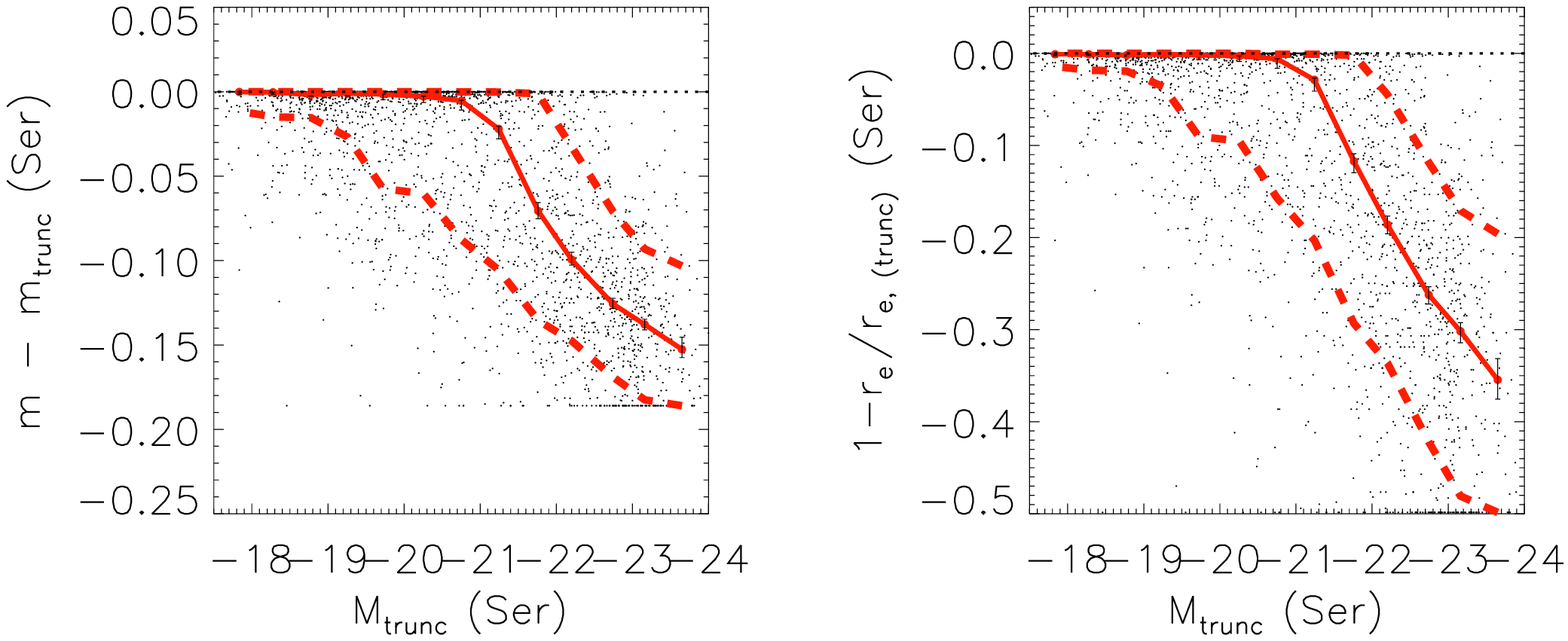}
  \includegraphics[width = 1.0\hsize]{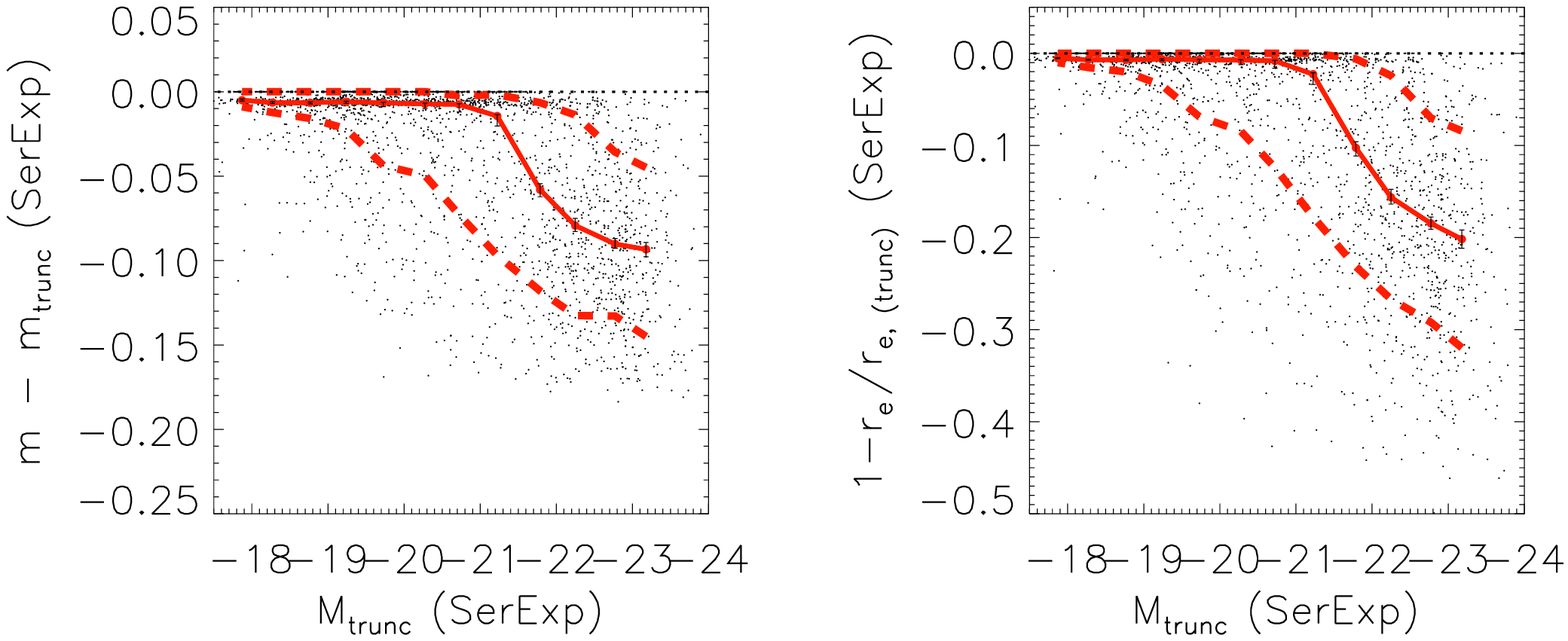}
  \vspace{-0.5cm}
  \caption{Top: Effect of truncation on the magnitude (left) and size (right) for the S{\'e}rsic fit as a function of absolute magnitude. Luminous galaxies tend to have larger $n$, so truncation matters more at high luminosity. Bottom: Similar to top panels but for the two-component SerExp fits (the S{\'e}rsic and Exponential profiles were truncated separately before combining them). In this and all following plots, the solid red line indicates the median of the data. The dashed red lines show the region which encloses 68\% of the galaxies at fixed absolute magnitude.}
 \label{trunceffser}
\end{figure}

\subsubsection{Truncation}
There is some discussion in the literature about what to report as the `total' light associated with a S{\'e}rsic profile. Whereas \cite{Meert2015} integrate their fits to infinity, others truncate the integral at approximately 7 or $8\times$ the fitted half-light radius (e.g. the SDSS pipeline). The radius which encloses half the truncated light is not usually reported. MPP-VAC provides both original and `truncated' values, which we now describe. 

Since PyMorph really performs 2D fits to images which are not usually circular, we truncate the light within elliptical isophotes. If $a_{e(\infty)}$ and $b_{e(\infty)}$ denote the lengths along the major and minor axes which include half the light before truncation, then we only include the light within an ellipse whose semi-major and semi-minor axes extend out to $7a_{e(\infty)}$ and $7b_{e(\infty)}$. (The combination $\sqrt{a_{e(\infty)} b_{e(\infty})}$ is sometimes called the effective radius $R_e$, so one might say that we truncate at $7R_e$.) In addition we also report the scale which encloses half of this truncated, rather than total, light. I.e., since we assume that truncation does not change the axis ratio, we report $b/a = b_{e(\infty)}/a_{e(\infty)}$, and the length of the semi-major axis which encloses half the truncated-light: $a_{e(\rm trunc)} < a_{e(\infty)}$. 

Figure~\ref{trunceff} shows the fractional changes to the total light and size which result from truncation: they are a deterministic function of S{\'e}rsic $n$. Since luminous galaxies tend to have larger $n$, truncation matters more at high luminosity. Since the $n-L$ correlation has scatter, one cannot simply translate the correction for $n$ into one for $L$. Therefore, Figure~\ref{trunceffser} shows the effect of truncation on the single-S{\'e}rsic light and size estimates of the galaxies in MPP-VAC (top panels). Similarly, the bottom panels show the effects for the two-component SerExp profiles (we truncate the S{\'e}rsic and Exponential profiles separately before combining them). These truncated magnitudes and sizes are also reported in MPP-VAC.

\begin{figure}
 \centering
 \includegraphics[width = 1.0\hsize]{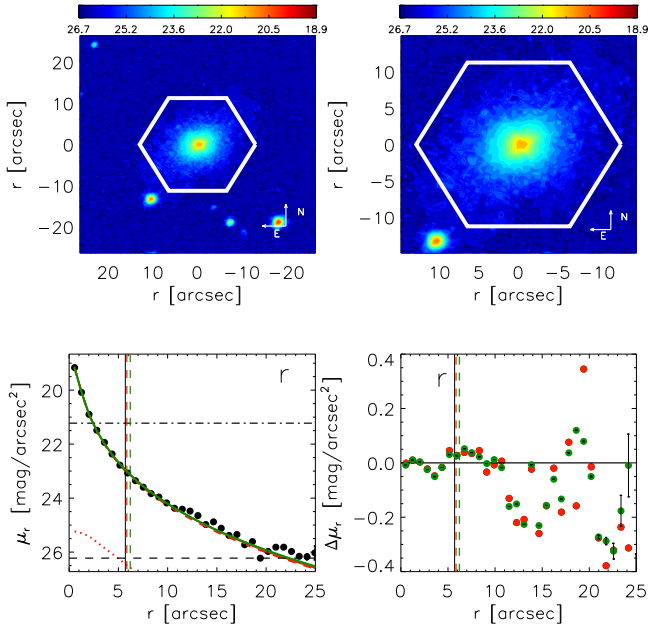}
 \caption{Example of a galaxy where a single-component fit is preferred (FLAG$\textunderscore$FIT$=1$) since the disk contribution is negligible. The different panels are similar to those in Figures~1-3.}
 \label{flag1}
\end{figure}

\begin{figure}
 \centering
 	 \includegraphics[width = 1.0\hsize]{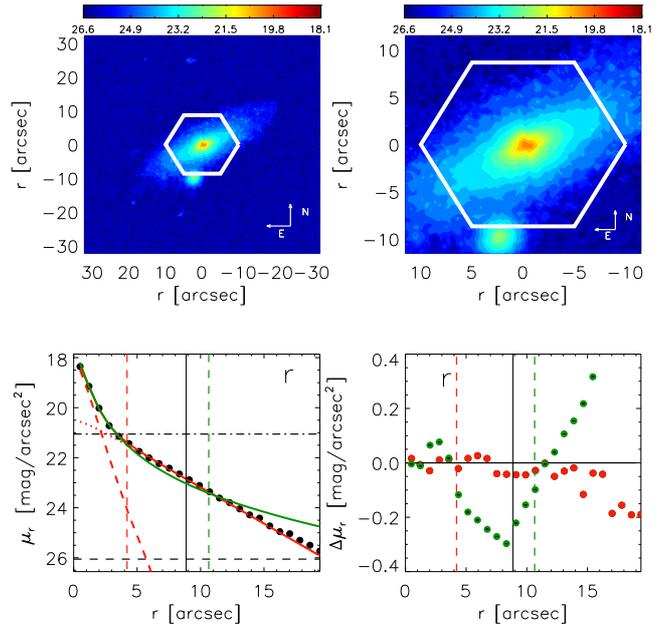}
\caption{Same as previous figure, except for this galaxy a two-component fit is preferred (FLAG$\textunderscore$FIT$=2$): the solid red curve provides a good fit over the entire image, whereas the green does not.}
 \label{flag2}
\end{figure}

\begin{figure}
 \centering
 	 \includegraphics[width = 1.0\hsize]{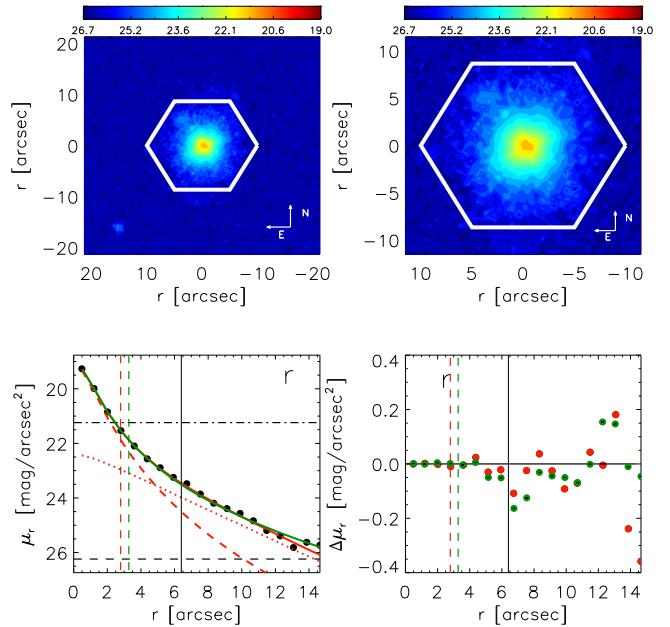}
\caption{Same as previous figure, except that for this galaxy the single-Ser and SerExp fits are both acceptable, so we do not express a preference for one over the other (FLAG$\textunderscore$FIT$=0$).}
 \label{flag0}
\end{figure}

\subsection{Description of the MPP-VAC catalog}\label{sec:cat}
The DR15 MaNGA release includes 4688 galaxy observations (identified by the PlateIFU and MaNGAID variables; some are repeated observations of the same galaxy). Of these 4688 observations, 16 are not in our MPP-VAC. These were either not galaxies (6), were too dim (3), or did not have a SDSS-DR14 identification and PSF (7). Thus MPP-VAC includes 4672 entries for 4599 unique  galaxies. Duplicate observations are defined with a match of 5 arcsec using the RA and DEC from the MaNGA datacubes (OBJRA, OBJDEC). We find 61 groups, i.e., there are 61 galaxies with multiple observations according to our criteria. (Note that this is slightly different from the number of unique MANGA-IDs: there are only 55 galaxies with repeated MANGA-ID.) Table~\ref{catcontents} shows the content of the catalog, which is in the FITS file format and includes 3 HDUs. Each HDU lists the parameters measured in the $g$, $r$, and $i$ bands, respectively. Table~\ref{catcontents} also provides three variables which identify galaxies with multiple MaNGA spectroscopic observations (see DUPL\_GR, DUPL\_N, and DUPL\_ID).

Note that PyMorph can have failures in its fitting. This is reported by the flags FLAG$\_$FAILED$\_$S and FLAG$\_$FAILED$\_$SE (it is set to 1 when we have a failed S{\'e}rsic or SerExp fit, respectively). For this catalog, 333 entries have FLAG$\_$FAILED$\_$S $=1$ and 406 have FLAG$\_$FAILED$\_$SE $=1$ in the $r$-band. There are 198 entries that have parameters from both PyMorph models that failed. Failures can happen for several reasons: contamination, peculiarity, bad-image, or bad model fit. Although, for the majority of this paper we use the set of parameters measured in the SDSS $r$-band, analysis from the other bands ($g$ and $i$) produce similar results (see Section~\ref{difbands}). The top part of Table~\ref{tabCat} lists the fraction of objects without photometric measurements for the different bands. About 8\% and 9\% of the objects do not have parameters from the S{\'e}rsic and SerExp fits, respectively. About 5\% of these objects do not have any PyMorph photometric parameters (i.e. FLAG$\textunderscore$FIT $=3$, see Section~\ref{flagfit}).




\subsubsection{Preference system: FLAG$\_$FIT}\label{flagfit}
We have introduced a flagging system, based on visual inspection, that indicates which fit is to be preferred for scientific analyses (see FLAG$\_$FIT in Table~\ref{catcontents}). This is because some galaxies -- Figure~\ref{flag1} shows an example -- are clearly just single-component objects. In Figure~\ref{flag1}, the disk component is irrelevant, so the associated disk parameters such as disk scale length are virtually meaningless. We set FLAG$\_$FIT$=1$ to indicate that the parameters from the single-S{\'e}rsic fit are preferred (even though $\chi^2_{\rm dof} = 1.041$ for both fits).

Other galaxies, such as the one in Figure~\ref{flag2}, are clearly made of two components, so we set FLAG$\_$FIT$=2$.  This is always supported by the goodness of fit:  in this $\chi^2_{\rm dof}=1.001$ for the two component SerExp fit, but 1.050 for the single component Ser fit. 

In some cases, the two models are both acceptable (e.g. Figure~\ref{flag0}, where $\chi^2_{\rm dof}=0.968$ and $0.971$ for the two- and single-component fits, respectively), so we set FLAG$\_$FIT$=0$.

MPP-VAC has 2367 entries with FLAG$\_$FIT$=$1, 1696 with FLAG$\_$FIT$=$2, and 411 FLAG$\_$FIT$=$0 in the $r$-band. The bottom part of Table~\ref{tabCat} lists the fraction of objects for each FLAG$\_$FIT type in the SDSS $g$, $r$, and $i$ bands.

We urge users to pay attention to the preferences expressed by FLAG$\_$FIT.

\subsection{Comparison with previous work}\label{sec:lit}

Compared to SDSS pipeline photometry, PyMorph fits return substantially more light for the most luminous galaxies \citep{Bernardi2013, Bernardi2017a, Bernardi2017b}. This is primarily because of differences in how the background sky is estimated and what model is fit to the surface brightness profile, although differences in the scale out to which one integrates the fit when defining the total luminosity also matter \citet[and references therein]{F2017}. 

In what follows, we compare the photometric parameters in the MPP-VAC, which we will refer to as DR15, with those from M15 as well as with analyses from two other groups. For single component fits, we compare with NSA photometry as well as with the S{\'e}rsic photometry of S11. For SerExp photometry, we compare with S11 only, as NSA do not provide two-component fits. 

\begin{table*}
\centering
MPP-VAC: The MaNGA PyMorph Photometric VAC
\begin{tabular}{ |p{3.5cm}|p{11.5cm}|p{2cm}| }
\hline 
Column Name & Description & Data Type \\ 
\hline 
IntID & Internal identification number & int \\ 
MANGA-ID & MaNGA identification  & string \\ 
PlateIFU & MaNGA PLATE-IFU & string \\ 
ObjID & SDSS-DR15 photometric identification number & long int \\ 
RA & Object right ascension (degrees) & double \\ 
Dec & Object declination (degrees) & double \\ 
z & NSA redshift or SDSS if NSA not available & float \\ 
extinction & SDSS extinction & float \\ 
DUPL\_GR & Group identification number for a galaxy with multiple MaNGA spectroscopic observations & int \\ 
DUPL\_N & Number of multiple MaNGA spectroscopic observations associated with DUPL\_GR & int \\ 
DUPL\_ID & Identification number of the galaxy in the group DUPL\_GR & int \\
FLAG\_FIT & Fit preference: No Preference (0), S{\'e}rsic (1), SerExp (2), S{\'e}rsic and SerExp failed (3) & int \\ 
FLAG\_FAILED\_S & This flag is set to 1 if the S{\'e}rsic fit failed (due to contamination/peculiarity/bad-image or bad model fit) otherwise is equal to 0 & int \\ 
M\_S & Total apparent magnitude from S{\'e}rsic fit & float \\ 
M\_S\_ERR & Error associated with M\_S & float \\ 
M\_S\_TRUNC & Truncated apparent magnitude to 7 $\times$ A\_hl\_S & float \\ 
A\_hl\_S & Half-light semi-major axis (arcsec) from S{\'e}rsic fit & float \\ 
A\_hl\_S\_ERR & Error associated with A\_hl\_S (arcsec) & float \\ 
A\_hl\_S\_TRUNC & Half-light semi-major axis (arcsec) associated with M\_TRUNC & float \\
N\_S & S{\'e}rsic index from S{\'e}rsic fit & float \\ 
N\_S\_ERR & Error associated with N\_S & float \\ 
BA\_S & Axis ratio (semi-minor/semi-major) from S{\'e}rsic fit & float \\
BA\_S\_ERR & Error associated with BA\_S & float \\ 
PA\_S & Position angle (degrees) from S{\'e}rsic fit & float \\ 
PA\_S\_ERR & Error associated with PA\_S (degrees) & float \\ 
GALSKY\_S & PyMorph sky brightness from S{\'e}rsic fit (mag/arcsec$^{2}$) & float \\ 
GALSKY\_S\_ERR &Error associated with GALSKY\_S (mag/arcsec$^{2}$) & float \\ 
FLAG\_FAILED\_SE & This flag is set to 1 if the SerExp fit failed (due to contamination/peculiarity/bad-image or bad model fit) otherwise is equal to 0 & int \\ 
M\_SE & Total apparent magnitude from SerExp fit & float \\
M\_SE\_TRUNC & Apparent magnitude from the truncated bulge and disk components of SerExp fit & float \\
A\_hl\_SE & Half-light semi-major axis (arcsec) of the total SerExp fit & float \\
A\_hl\_SE\_TRUNC & Half-light semi-major axis (arcsec) associated with M\_SE\_TRUNC & float \\
BA\_SE & Axis ratio (semi-minor/semi-major) of the total SerExp fit & float \\
BT\_SE & B/T (bulge-to-total light ratio) from SerExp fit & float \\
BT\_TRUNC & B/T from the truncated bulge and disk components of SerExp fit & float \\
M\_SE\_BULGE & Bulge apparent magnitude from SerExp fit & float \\
M\_SE\_BULGE\_ERR & Error associated with M\_SE\_BULGE & float \\
M\_SE\_BULGE\_TRUNC & Bulge apparent magnitude truncated to 7 $\times$ A\_hl\_SE\_BULGE  & float \\
A\_hl\_SE\_BULGE & Bulge half-light semi-major axis (arcsec) from SerExp fit  & float \\
A\_hl\_SE\_BULGE\_ERR & Error associated with A\_hl\_SE\_BULGE (arcsec) & float \\
A\_hl\_SE\_BULGE\_TRUNC & Bulge half-light semi-major axis (arcsec) associated with M\_SE\_BULGE\_TRUNC & float \\
N\_SE\_BULGE & Bulge S{\'e}rsic index from SerExp fit (galaxies with flipped components have N\_SE\_BULGE = 1 AND N\_SE\_DISK $\leq$ 1) & float \\
N\_SE\_BULGE\_ERR & Error associated with N\_SE\_BULGE & float \\
BA\_SE\_BULGE & Bulge axis ratio (semi-minor/semi-major) from SerExp fit & float \\
BA\_BULGE\_ERR & Error associated with BA\_SE\_BULGE  & float \\
PA\_SE\_BULGE & Bulge position angle (degrees) from SerExp fit  & float \\
PA\_SE\_BULGE\_ERR & Error associated with PA\_SE\_BULGE (degrees) & float \\
M\_SE\_DISK & Disk apparent magnitude from SerExp fit & float \\
M\_SE\_DISK\_ERR & Error associated with M\_SE\_DISK  & float \\
M\_SE\_DISK\_TRUNC & Disk apparent magnitude truncated to 7 $\times$ A\_hl\_SE\_Disk & float \\
A\_hl\_SE\_DISK & Disk half-light semi-major axis (arcsec) from SerExp fit (Note: it is not the disk scale) & float \\
A\_hl\_SE\_DISK\_ERR & Error associated with A\_hl\_SE\_DISK (arcsec) & float \\
A\_hl\_SE\_DISK\_TRUNC & Disk half-light semi-major axis (arcsec) associated with M\_SE\_DISK\_TRUNC & float \\
N\_SE\_DISK & Disk S{\'e}rsic index from SerExp fit (galaxies with flipped components have N\_SE\_BULGE = 1 and N\_SE\_DISK $\leq$ 1) & float \\
N\_SE\_DISK\_ERR & Error associated with N\_SE\_DISK & float \\
BA\_SE\_DISK & Disk axis ratio (semi-minor/semi-major) from SerExp fit & float \\
BA\_SE\_DISK\_ERR & Error associated with BA\_SE\_DISK & float \\
PA\_SE\_DISK & Disk position angle (degrees) from SerExp fit & float \\
PA\_SE\_DISK\_ERR & Error associated with PA\_SE\_DISK (degrees) & float \\
GALSKY\_SE & PyMorph sky brightness (mag/arcsec$^{2}$) from SerExp fit & float \\
\end{tabular}
\caption{The photometric parameters listed in this catalog were obtained from S{\'e}rsic and S{\'e}rsic+Exponential fits to the SDSS images from the latest processing reduction, i.e. post-DR12. The table includes three data extensions for the $g$, $r$, and $i$ bands.  Note that all position angles here are with respect to the camera columns in the SDSS ``fpC'' images (which are not aligned with the North direction); to convert to the convention where North is up, East is left set PA$_{\rm MaNGA}$ = (90-PA$_{\rm PyMorph}$) - SPA, where PA$_{\rm PyMorph}$ is the value given in this Table, and SPA is the SDSS camera column position angle with respect to North reported in the primary header of the ``fpC'' SDSS images. PA$_{\rm MaNGA}$ increases from East towards North. The SPA angles for this catalog are provided in a separate file which can be downloaded from the same MPP-VAC website$^3$.} 
\label{Table1}
\label{catcontents}
\end{table*}

\begin{table}
\centering
FRACTION OF GALAXIES\\
\begin{tabular}{cccc}
  \hline
 Band & FLAG$\_$FAILED$\_$S & FLAG$\textunderscore$FAILED$\textunderscore$SE & FLAG$\textunderscore$FIT  \\
 & = 1 & = 1 & = 3 \\
\hline 
       $g$  &  0.075  &   0.089 &   0.046 \\
       $r$  &  0.071  &  0.087  &  0.042  \\
       $i$  &  0.085 &   0.093  &  0.045 \\ 
 \hline
 \hline
 \end{tabular}
 Galaxies with FLAG$\textunderscore$FIT $\ne$ 3\\
 \begin{tabular}{cccc}
 \hline 
Band & FLAG$\textunderscore$FIT & FLAG$\textunderscore$FIT & FLAG$\textunderscore$FIT  \\
 & = 0 & = 1 & = 2 \\
 \hline
       $g$ &   0.087 &   0.526 &  0.341 \\
       $r$ &   0.088  &   0.507  & 0.363 \\
       $i$ &   0.087 &  0.506 &  0.362 \\
 \hline
 \hline
\end{tabular}
\caption{Top part: Fraction of galaxies which do not have PyMorph parameters from S{\'e}rsic (FLAG$\_$FAILED$\_$S = 1), SerExp (FLAG$\_$FAILED$\_$SE = 1) or both (FLAG$\textunderscore$FIT = 3) in the SDSS $g$, $r$ and $i$ bands. Bottom part: Fraction of galaxies which have either S{\'e}rsic or SerExp or both set of parameters and flagged as having 2 components (FLAG$\textunderscore$FIT = 2), 1 component (FLAG$\textunderscore$FIT = 1), or for which both descriptions are equally acceptable (FLAG$\textunderscore$FIT = 0).}
\label{tabCat}
\end{table}

\begin{figure}
 \centering
 	 \includegraphics[width = 1.0\hsize]{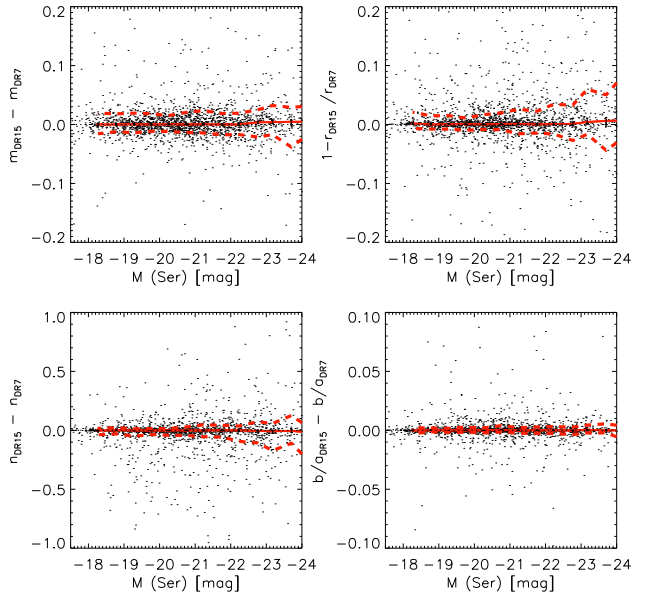}
	  \vspace{-0.5cm}
\caption{Comparison of PyMorph S{\'e}rsic photometric parameters in DR7 \citep{Meert2015} and those in MPP-VAC (subscript DR15 in the figure) for galaxies with FLAG$\textunderscore$FIT $=0$ or $=1$. The figure shows that the difference in apparent magnitude, size, S{\'e}rsic index and axis ratio is small. The solid red line indicates the median of the data. The dashed red lines show the region which encloses 68\% of the galaxies at fixed absolute magnitude.}
 \label{catAlans}
\end{figure}
	  
\begin{figure}
 \centering	  
 	 \includegraphics[width = 1.0\hsize]{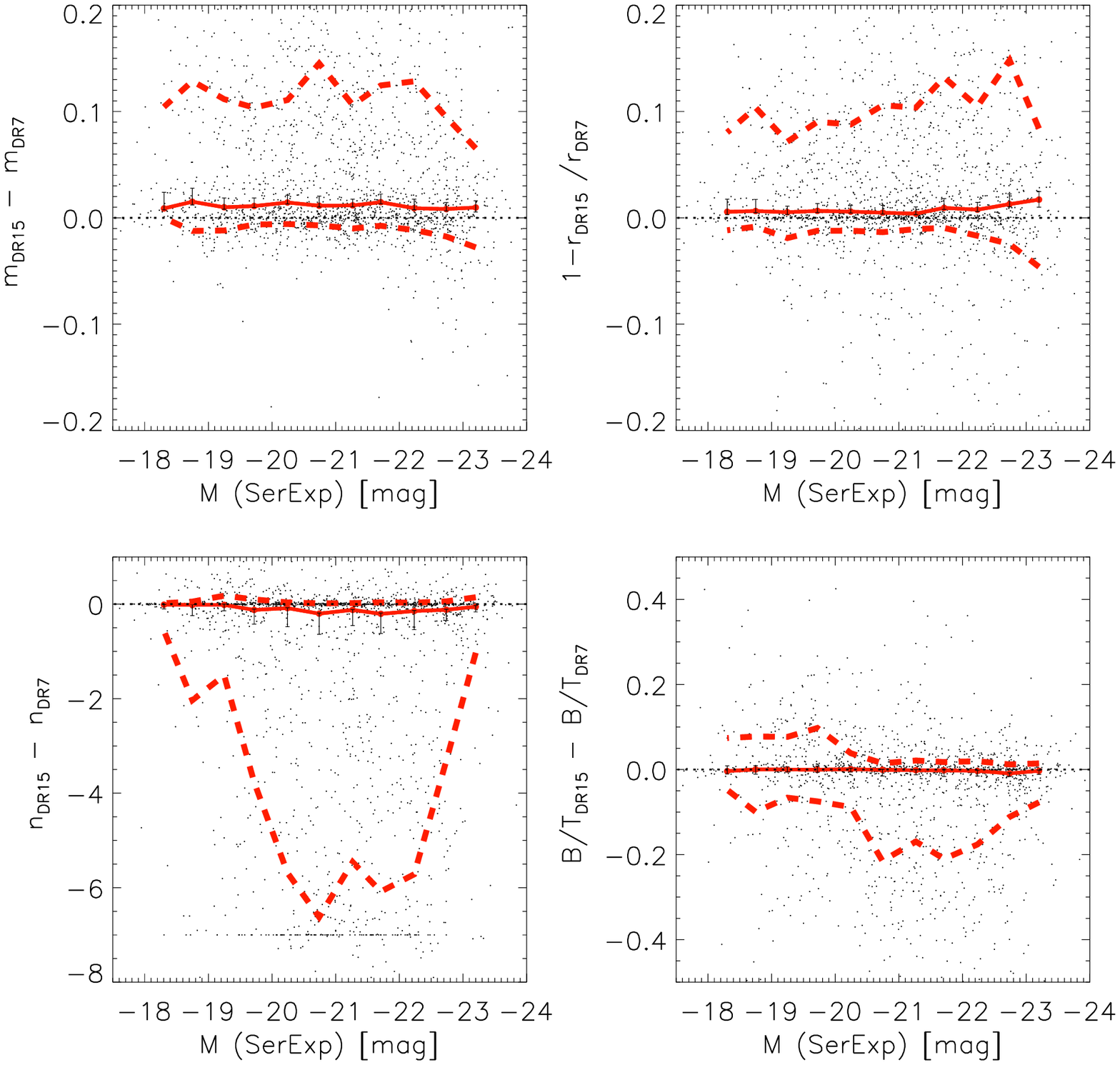}
	   \vspace{-0.5cm}
\caption{Same as previous figure, but now for PyMorph SerExp fits for galaxies with FLAG$\textunderscore$FIT $=0$ or $=2$. For most objects, the agreement is again very good. However, the asymmetric scatter around the median is due to a cloud of outliers associated with our flipping and/or re-fitting, for which our DR15 analysis returns fainter magnitudes, smaller sizes, smaller $n$-bulge, and smaller B/T ratios. The effect is more evident for the S{\'e}rsic index $n$ comparison (bottom left panel) since our DR15 analysis has several more galaxies with $n=1$ but many fewer $n=8$ compared to DR7.}
 \label{catAlanse}
\end{figure}




\subsubsection{Previous PyMorph analyses: Meert et al. 2015 (DR7)}

The most straightforward comparison is with the analysis of SDSS DR7 images of M15, which includes about 85\% of the MPP-VAC objects. Since both use PyMorph (but different SDSS images processing reduction -- DR7 versus post-DR12), we expect little difference for S{\'e}rsic photometry, with more substantial changes due to refitting of the SerExp photometry. Figure~\ref{catAlans} shows that, indeed, for S{\'e}rsic photometry, the changes in apparent magnitude, size, S{\'e}rsic index, and axis ratio are all small ($rms$ scatter of a few \%). Figure~\ref{catAlanse} shows that they are also very similar for SerExp photometry, except for the cloud of outliers associated with our refitting and/or flipping. For these outliers, our DR15 analysis returns fainter magnitudes, smaller sizes, smaller $n$, and smaller B/T ratios. The effect is more evident for the S{\'e}rsic index $n$ comparison (bottom left panel) since our DR15 analysis has several more galaxies with $n=1$ but many fewer $n=8$ compared to DR7.

\begin{figure*}
 \centering
 \includegraphics[width = 0.9\hsize]{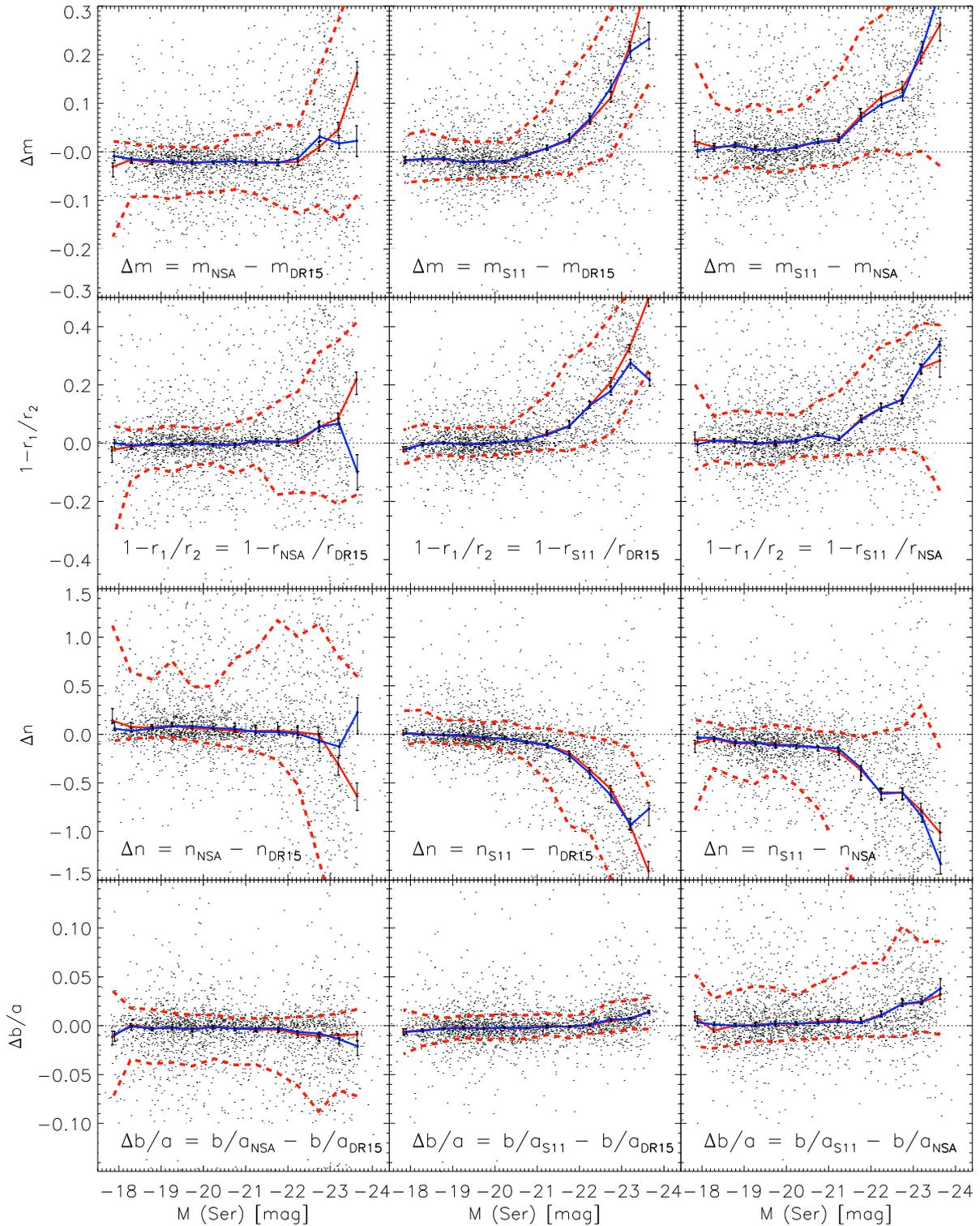}
   \vspace{-1.cm}
   \caption{Comparison of PyMorph DR15 single-S{\'e}rsic magnitudes (top row), sizes (second row from top), S{\'e}rsic indices (third row), and axis ratio $b/a$ (bottom row) with corresponding values from NSA and S11 for galaxies with FLAG$\textunderscore$FIT $=0$ or $1$.
Solid lines indicate the median of the data. The dashed lines show the region which encloses 68\% of the galaxies at fixed absolute magnitude. In all cases, we show results as a function of PyMorph magnitude (red lines); using NSA magnitude instead (blue lines) makes little difference except in the top middle panel, in which the trend with $\rm M$ is even larger. In all cases, PyMorph and NSA are in good agreement (left), whereas offsets between PyMorph and S11 (middle) are like those between NSA and S11 (right).}
 \label{SerMag}
\end{figure*}

\begin{figure}
 \centering
 	 \includegraphics[width = 1.\hsize]{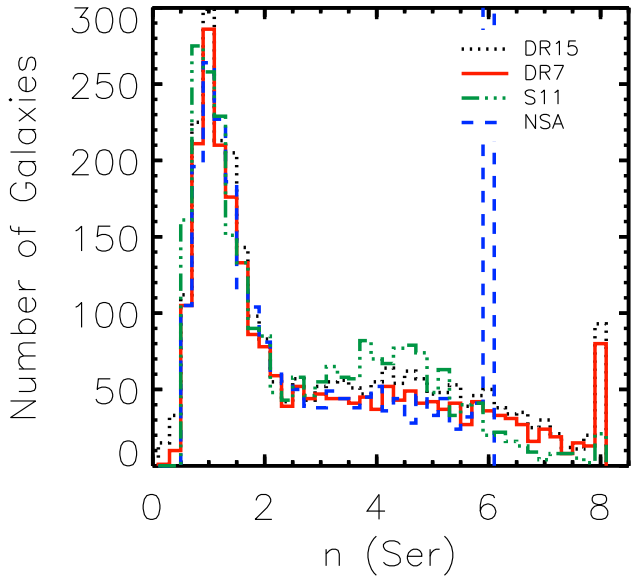}
	  \vspace{-0.5cm}
          \caption{Distribution of S{\'e}rsic $n$ for galaxies with FLAG$\textunderscore$FIT $=0$ or $=1$ (i.e. single-component fit is preferred). Black histogram is PyMorph DR15; red, green, and blue show DR7 \citep{Meert2015}, S11, and NSA, respectively. Our DR15 analysis limits $n \le 8$, whereas the S11 analysis allows $0.5 \le n \le 8$, and NSA does not allow $n>6$. This explains the spike at $n=6$ where NSA has 697 galaxies.}
 \label{Sernhist}
\end{figure}

\begin{figure}
 \centering
 	\includegraphics[width = 1.0\hsize]{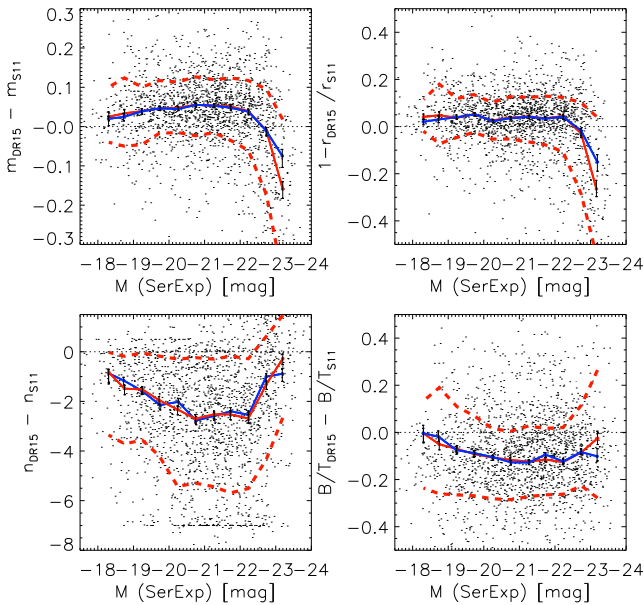}
	 \vspace{-0.5cm}
          \caption{Comparison of PyMorph DR15 two-component SerExp parameters with S11 for galaxies with FLAG$\textunderscore$FIT $=0$ or $=2$. From left to right, top to bottom: apparent total magnitude, half-light semi-major axis of the total SerExp fit, bulge S{\'e}rsic index, and B/T. We show results as a function of PyMorph absolute magnitude (red lines) and using S11 absolute magnitudes (blue lines).}
 \label{litSerExp}
\end{figure}

\begin{figure}
 \centering
 	 \includegraphics[width = 1.0\hsize]{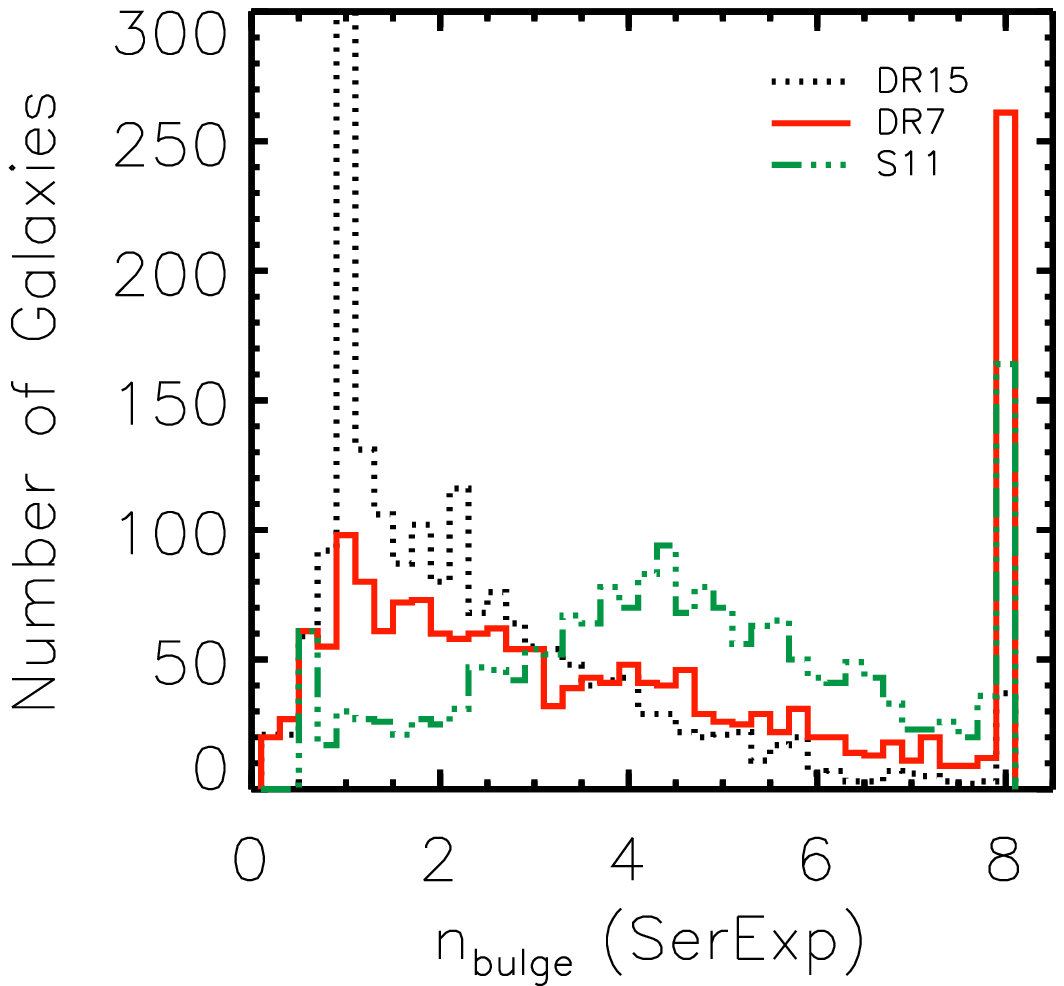}
	  \vspace{-0.5cm}
          \caption{Distribution of S{\'e}rsic index $n$ for the bulge components of galaxies with FLAG$\textunderscore$FIT $=0$ or $=2$ (i.e. two-component SerExp fit is preferred). Black histogram is PyMorph DR15; red and green show DR7 (M15) and S11. Our DR15 analysis has several more galaxies with $n=1$ but many fewer $n=8$ compared to the DR7 analysis, as a result of our eye-ball motived refitting and flipping. The spike at $n=1$ for DR15 extends to 529 galaxies. See text for discussion of why the S11 distribution shows a peak at $n\sim 4$. } 
 \label{SerExpnhist}
\end{figure}

\subsubsection{Non-PyMorph single-S{\'e}rsic fits}
We now compare with S11 who also provide S{\'e}rsic-photometry for galaxies within the Legacy area of the SDSS DR7. For the S{\'e}rsic comparison, we have $\sim 94$\% of the galaxies in common with FLAG$\textunderscore$FIT$=0$ or FLAG$\textunderscore$FIT$=1$. 

M15 and \cite{F2017} showed that the S11 analysis is slightly biased because it used an overestimate of the background sky, and so tends to underestimate the light of the most luminous or most extended galaxies.

We also compare with photometry from the NSA catalog, where we have $\sim 98$\% of the galaxies in common, in which issues with the sky have been resolved (following \citealt{Blanton2011}; see \citealt{F2017} for further discussion). 

Figure~\ref{SerMag} compares our PyMorph DR15 single-S{\'e}rsic magnitudes, sizes, S{\'e}rsic indices, and axis ratios $b/a$ with NSA and S11 for galaxies with FLAG$\textunderscore$FIT $=0$ or $=1$. The top left panel shows that PyMorph is about 0.02~mags fainter than NSA, but it is otherwise in good agreement. However, it can be more than 0.1~mags brighter than S11 for the most luminous galaxies (top middle). This is because S11 measurements are biased by an overestimate of the background sky. Indeed, a comparison of S11 with NSA magnitudes shows a similar trend (top right panel).
The second row (from top) of Figure~\ref{SerMag} shows a similar comparison of the S{\'e}rsic half-light size estimates. Again, our PyMorph DR15 estimates are in good agreement with NSA, whereas S11 sizes are biased to smaller sizes, consistent with the fact that S11 assumed a brighter background sky. The third row (from top) shows that PyMorph and NSA return similar estimates of $n$, except possibly for the most luminous objects. Some of this is because our DR15 analysis allows $n \le 8$, whereas NSA only allows $n \le 6$ (and S11 require $0.5\le n\le 8$). However, S11 tends to be systematically smaller than both, especially for the most luminous objects. Finally, the bottom row shows that estimates of the axis ratio are in very good agreement between the different work.

Figure~\ref{Sernhist} shows the distribution of the S{\'e}rsic index $n$. While DR15, S11, and NSA all show a similar concentration of values around $n=1$, with a long tail to longer $n$, it appears that S11 tends to favor $n\sim 4$ slightly compared to DR15 or NSA. The spikes in the distribution of the S{\'e}rsic index at $n=6$ and $n=8$ are due to the limits in $n$ imposed by the different groups.

\subsubsection{Non-PyMorph two-component SerExp fits}
We now perform a similar comparison of our PyMorph DR15 SerExp photometry with previous non-PyMorph work. This is only possible with S11, as the NSA catalog only reports parameters from single-S{\'e}rsic fits. For this comparison of the SerExp fits, we have 1931 galaxies in common with S11. In view of our eyeball-based re-fitting and flipping, which we described in Section~\ref{sec:refit} and ~\ref{sec:flip}, we are expecting much larger differences here than we found for the single-S{\'e}rsic photometry. 

Figure~\ref{litSerExp} shows that our DR15 SerExp photometry returns slightly less light, smaller sizes, smaller $n$-bulge, and smaller B/T. Most of these differences are driven by the relatively large offset in $n$. 

Figure~\ref{SerExpnhist} compares our DR15 bulge S{\'e}rsic indices with those from the M15 PyMorph analysis of DR7, and from S11. For S11, the distribution peaks around $n_{\rm bulge}=4$; this may be because, in cases where $n_{\rm bulge}$ is not well constrained, S11 returns the median of the allowed prior range $0.5\le n\le 8$. 

\begin{table*}
\centering
MDLM-VAC: The MaNGA Deep Learning Morphological VAC
\begin{tabular}{ |p{3.5cm}|p{11.5cm}|p{2cm}| }
\hline 
Column Name & Description & Data Type \\ 
\hline 
IntID & Internal identification number & int \\ 
MANGA-ID & MaNGA identification  & string \\ 
PlateIFU & MaNGA PLATE-IFU & string \\ 
ObjID & SDSS-DR15 photometric identification number & long int \\ 
RA & Object right ascension (degrees) & double \\ 
Dec & Object declination (degrees) & double \\ 
z & NSA redshift & float \\ 
DUPL\_GR & Group identification number for galaxies with multiple MaNGA observations & int \\ 
DUPL\_N & Number of multiple MaNGA observations associated with DUPL\_GR & int \\ 
DUPL\_ID & Identification number of the galaxy in the group DUPL\_GR & int \\
TType & TType value. TType$ <$ 0 for ``early-type" galaxies. TType$>$ 0 for ``late-type" galaxies & double \\ 
flag\_TT & This value indicates if the TType has been changed after a visual inspection (0=no, 1=yes) & int \\ 
P\_S0 & Probability of being S0 rather than E. Only meaningful for galaxies with TType $\leq$ 0 & double \\ 
flag\_S0 & This value indicates if the P\_S0 has been changed after a visual inspection (0=no, 1=yes) & int \\ 
P\_edge\_on & Probability of being edge-on & double \\ 
P\_bar\_GZ2 & Probability of having a bar signature (trained with GZ2 catalog). Edge-on galaxies should be removed to avoid contamination & double\\ 
P\_bar\_N10 & Probability of having a bar signature (trained with N10 catalog). No contaminated by edge-on galaxies & double \\ 
P\_merg & Probability of merger signature (or projected pair) & double \\
P\_bulge & Probability of having a dominant bulge vs. no bulge & double \\ 
P\_cigar & Probability of having cigar shape vs. round shape & double \\ 
\hline 
\hline 
\end{tabular}
\caption{Content of the Deep Learning morphological catalog for the DR15 MaNGA sample. This catalog is available online$^4$.} 
\label{morphcatcontents}
\end{table*} 

This peak is not present in either of the PyMorph analyses. Our DR15 analysis has several more $n=1$ but many fewer $n=8$ compared to the DR7 analysis, as a result of our eye-ball motived re-fitting and flipping. Of course, this also affects B/T, but we reserve this comparison for the next section.
 

\section{MaNGA Deep Learning Morphology Value Added catalog (MDLM-VAC)}\label{sec:morphcat}

Morphological classifications are available for all the objects in the MPP-VAC. These are provided in the MaNGA Deep Learning Morphology Value Added catalog (MDLM-VAC) which is available online\footnote{www.sdss.org/dr15/data\textunderscore access/value-added-catalogs/manga-morphology-deep-learning-dr15-catalog}. In this section we describe the MDLM-VAC and explain how it was constructed. 

\subsection{Catalog content and description}\label{sec:morphcatalog}

The MDLM-VAC contains Deep Learning (DL) based morphological classifications for the same sample as the MPP-VAC. The methodology for training and testing the DL models is described in detail in \citet[hereafter DS18]{DS2018}, where classifications for about 670,000 objects from the SDSS DR7 Main Galaxy Sample of \citealt{Meert2015} are provided. Since about 15\% of the MaNGA DR15 galaxies were not included in that analysis, the present catalog provides a homogenous morphological catalog for all of the MaNGA DR15 sample. We strongly recommend reading DS18 for a better understanding of the catalog construction, meaning, and usage. 

In short, the DL morphologies are obtained by training a Convolutional Neural Network with two visually-based morphological catalogs: \cite{Willett2013} and \cite{Nair2010}. The algorithm takes as input RGB SDSS-DR7 images in \textit{.jpg} format. We train one model for each classification task. The training is an iterative process which determines a set of weights that minimizes the difference between the input classification and the DL model output. Once the weights have been optimized, the DL algorithm applies them to new galaxy images not used in the training, providing a classification for each of them.

The DL algorithm was trained and tested with SDSS-DR7 cutouts, so we can easily apply the models to the DR7 images of the MaNGA DR15 galaxies. The time required for classifying a new set of $\sim$ 5000 galaxies once the models are trained is minimal (minutes). This means that morphological classification for future MaNGA data releases will be available essentially as soon as the data are made public. The performance of the models in this new dataset, in terms of accuracy, completeness, and contamination, should be comparable to the results in DS18 ($>$ 90\% for all tasks). The values contained in this catalog may be slightly different from the ones given in DS18 (for the galaxies in common) due to small variations in centering or cutout size, which, for this sample, are based on SDSS DR15 instead of DR7.

Table~\ref{morphcatcontents} shows the format of the MDLM-VAC. It provides parameters obtained by applying DL models trained with the \citealt{Nair2010} catalog: a TType value, a finer separation between S0 and pure ellipticals (E), and the probability of having a bar feature. All the additional set of morphological properties are obtained by applying DL models trained with the Galaxy Zoo 2 catalog (\citealt{Willett2013}; hereafter GZ2). The DL models are trained in binary mode, so the output is the probability that a galaxy belongs to the stated class (e.g., $P_{edge-on}$ is the probability that a galaxy is edge-on): these probabilities take values in the range [0, 1].  Since the models return probabilities, a user-defined threshold value ($P_{\rm thr}$) can be used to select objects of a certain type. With this in mind, values of precision ($\sim$ purity) and True Positive Rate (TPR $\sim$ completeness) for three $P_{\rm thr}$ values are tabulated in Table 2 of DS18.

The TType model is instead trained in regression mode, so the output is directly the TType of each galaxy, with values ranging from [-3,10]. The typical error in TType is $\sim$1.1. See Figure 13 of DS18 for a better understanding of the TType values presented in this catalogue, as well as for $P_{S0}$ (which is only meaningful for ``early-type" galaxies, i.e., when TType $\le$ 0).  
  
While MDLM-VAC obviously complements the parameters in MPP-VAC, it also complements the available estimates from GZ2 by providing a TType and a finer separation between S0s and pure ellipticals. For the parameters in common with the GZ2 ($P_{edge-on}$, $P_{bar}$, $P_{bulge}$, $P_{cigar}$), the DL-output probability distributions are more bimodal, reducing the fraction of galaxies with an uncertain classification (see discussion in DS18).  See Section~\ref{sec:gz} for a more detailed comparison between MDLM-VAC TType and GZ2 parameters reported by \cite{Willett2013}.

 Given the reasonable size of the sample, all the TType and $P_{S0}$ values have been eye-balled for additional reliability. A flag is provided for the TType and $P_{S0}$, indicating when the original output of the model has been changed after visual inspection. This was only necessary for a small fraction of the objects in our sample:  We changed TType for less than 3\% of the objects, and modified $P_{S0}$ for about 5\% of the objects with TType$\leq 0$. Most mis-classifications are due to incorrect radius values (used for the cutout size), faint galaxies, or contamination by nearby objects.

We remark that the $P_{\rm merg}$ value is a good indicator of projected pairs or nearby objects rather than of real on-going mergers. We found it extremely useful for identifying galaxies whose MaNGA spectroscopic data was contaminated by neighbors. We find that $\sim$ 50\% of the galaxies with a contaminated spectrum have $P_{\rm merg}$ $>$ 0.5, compared to $\sim$ 17\% for the whole sample. Increasing the limit, only $\sim$ 11\% of the whole sample has $P_{\rm merg}$ $>$ 0.8, while this fraction is $\sim$ 40\% for the contaminated sample.

\subsection{Our morphological classifications}\label{sec:morph}

\begin{table}
\centering
FRACTION OF GALAXIES\\
\begin{tabular}{lccc}
  \hline
  Type & FLAG$\textunderscore$FIT & NoContam & Both+$\sigma_0 > 0$ \\
  & $\ne 3$ &   &   \\
  \hline
  E & 0.952 & 0.772 & 0.772 \\
  S0 & 0.968 & 0.907 & 0.853 \\
  0 $<$ TType $<$ 3 & 0.971 & 0.884 & 0.842 \\
  TType $>$ 3 & 0.949 & 0.878 & 0.812 \\
  \hline
  \hline
\end{tabular}
\\
  Galaxies with FLAG$\textunderscore$FIT $\ne 3$+NoContam+$\sigma_0 > 0$ 
\begin{tabular}{lccc}
  \hline
  Type & FLAG$\textunderscore$FIT & FLAG$\textunderscore$FIT & FLAG$\textunderscore$FIT  \\
  & = 0 & = 1 & = 2 \\
 \hline
 E & 0.234 & 0.587 & 0.179  \\
 S0 & 0.163 & 0.419 & 0.419 \\
 0 $<$ TType $<$ 3 & 0.021 & 0.463 & 0.516  \\
 TType $>$ 3 & 0.003 & 0.576 & 0.421  \\
 \hline
 \hline
\end{tabular}
\caption{Top part: Fraction of galaxies of a given morphological type which have PyMorph parameters (from S{\'e}rsic and/or SerExp), good spectra (no contamination), and with central velocity dispersion $\sigma_0>0$. Bottom part: Fraction of galaxies which satisfy all criteria reported in the top part of the table and flagged as having 2 components (FLAG$\textunderscore$FIT = 2), 1 component (FLAG$\textunderscore$FIT = 1), or for which both descriptions are equally acceptable (FLAG$\textunderscore$FIT = 0).}
\label{tabfit}
\end{table}

\begin{figure}
 \centering
  \includegraphics[width = 1.1\hsize]{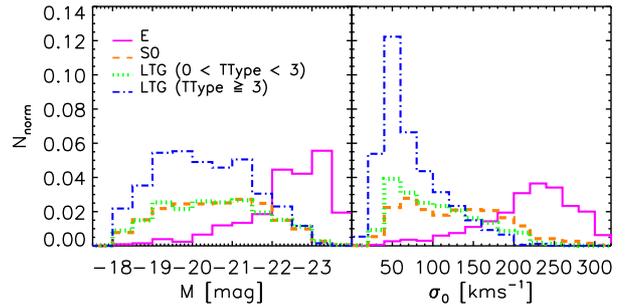}
  \vspace{-0.5cm}
  \caption{Contribution to the distribution of luminosities (left) and central velocity dispersions (right) from different morphological types (as labeled). Es dominate at large luminosities and $\sigma_0$, whereas LTGs with TType $>3$ dominate at small $\sigma_0$.}
 \label{morphMS}
\end{figure}

\begin{figure}
 \centering
  \includegraphics[width = 1.\hsize]{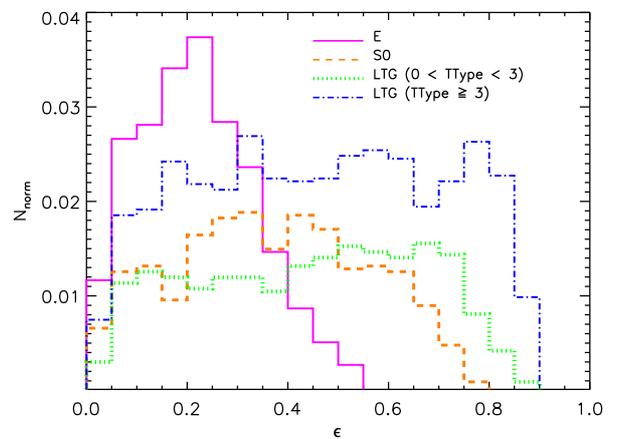}
  \vspace{-0.5cm}
  \caption{Distribution of $\epsilon = 1-b/a$ for different morphological types (as labeled).  Es are well-peaked around $\epsilon=0.2$, whereas LTGs with TType$\ge 3$ are approximately uniformly distributed over the entire range.}
 \label{morphba}
\end{figure}

In the analyses which follow, we mainly use the TType and $P_{\rm S0}$ from MDLM-VAC to separate objects into classes (as suggested by DS18). Specifically we use them to define two classes of late-type galaxies (LTG):
those with TType $> 3$, and others with $0<$ TType $<3$; 
objects having TType $< 0$ and P$_{\rm S0} \ge 0.5$ are defined to be S0; and ellipticals (E) have TType $< 0$ and P$_{\rm S0} < 0.5$.
Figure~14 in DS18 shows that essentially all Es have $P_{\rm S0}<0.5$, so it is very unlikely that our sample of S0s is contaminated by Es.  On the other hand, Figure~12 in DS18 shows that while TType$=0$ is a reasonable choice for separating S0s from other LTGs, the actual boundary is not particularly sharp.  By setting the threshold at TType$=0$, objects we classify as LTGs may be contaminated by S0s, more than vice versa.  We have also compared our morphological classification with the GZ2 parameters P$_{\rm Smooth}$ and P$_{\rm Disk}$ that are commonly used to define ``early-type" and ``late-type" galaxies (e.g., \citealt{parikh2018, Lee2018}). As we discuss in Section~\ref{sec:gz}, we believe that our classification based on the above criteria is superior to that provided by the GZ2 based on P$_{\rm Smooth}$ or P$_{\rm Disk}$.

\section{Photometry and morphology}\label{sec:photmorph}

In this section, we consider some illustrative science which results from combining the MPP-VAC with the MDLM-VAC.  

The second column in the top part of Table~\ref{tabfit} lists the fraction of objects associated with each morphological classification with reliable PyMorph estimates (recall from Section~\ref{sec:cat} that FLAG$\textunderscore$FIT $=3$ flags objects for which PyMorph failed).  However, not all of these have uncontaminated `deblended' spectra:  the third column lists the fraction of objects with both FLAG$\textunderscore$FIT $\ne 3$ and uncontaminated spectra. Sometimes for these objects, a reliable estimate of the central velocity dispersion $\sigma_0$ is not available.  Since we need $\sigma_0$ in what follows, we only work with objects having FLAG$\textunderscore$FIT $\ne 3$ and uncontaminated spectra and reliable $\sigma_0$.  The final column of the Table lists the fraction of objects which satisfy all three criteria.
The bottom part of the Table shows the fraction of these objects which are flagged as having 2 components, 1 component, or for which both descriptions are equally acceptable.

\begin{figure}
 \centering
\includegraphics[width = 1.1\hsize]{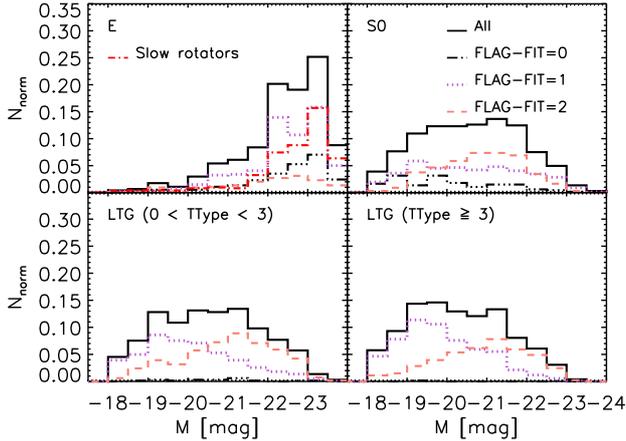}
  \vspace{-0.75cm}
  \caption{Distribution of luminosities for objects classified as being Es, S0s, and LTGs with TType smaller and greater than 3 (top left, top right, bottom left and right). Dotted and dashed histograms in each panel show objects classified as being composed of 1 or 2 components (FLAG$\textunderscore$FIT = 1 and 2, respectively), while dot-dot-dashed histogram shows the distribution of galaxies for which both fits are equally acceptable (FLAG$\textunderscore$FIT = 0). Red dot-dashed histogram in top left panel shows the Es that are `slow rotators' ($\sim 60\%$ of Es). In the bottom panels, two-component systems (dashed) tend to be more luminous.}
 \label{morphM}
\end{figure}

Figure~\ref{morphMS} shows the distribution of luminosity and central velocity dispersion $\sigma_0$ for these objects, subdivided by morphological type. (These are not luminosity and velocity dispersion functions in the usual sense, because we have not accounted for MaNGA's selection procedure.) Es dominate the counts at large luminosities and $\sigma_0$; S0s or LTGs with TType $<3$ tend to be similar to one another, and tend to have smaller $L$ and $\sigma_0$ than Es; and LTGs with TType $>3$ dominate at small $\sigma_0$. Note that neither $\sigma_0$ nor PyMorph photometry played any role in the morphological classification.

Figure~\ref{morphba} shows a similar study of the distribution of $\epsilon = 1-b/a$.  For objects having FLAG$\textunderscore$FIT=0 or FLAG$\textunderscore$FIT=2, the quantity $b/a$ is the semi-minor/semi-major axis ratio of the total SerExp fit (i.e. BA\_SE in Table~\ref{catcontents}). Again, there is a nice correlation with morphology, even though PyMorph $b/a$ played no role in the classification.  Es have a narrow distribution which peaks around $\epsilon \sim 0.2$ (the decrease at large $\epsilon$ is due to the lack of a disk/rotational component, while the decrease at low $\epsilon$ is expected to be due to triaxiality; see e.g. \citealt{Lambas1992}); S0s have a broader distribution than Es, but they do not extend beyond about 0.7; LTGs with $0\le$TType$\le 3$ extend to about 0.8; and LTGs with TType$\ge 3$ have a uniform distribution over almost the entire range (the decrease at large $\epsilon$ is due to the presence of a small bulge and/or the fact that the disk is not infinitely thin; while at low $\epsilon$ this is probably due to triaxiality). The differences between Es and LTGs are rather similar to those based on Galaxy Zoo classifications by \cite{Rodriguez2013}. These trends are also consistent with \cite{Lambas1992}, except for S0s, for which we find a broader distribution.  On the other hand, S0s account for many of the `fast rotators' in the top panel of Figure~5 of \cite{Weijmans2014}; these span a broad range of $\epsilon$, consistent with our Figure~\ref{morphba}.

\begin{figure}
 \centering
  \includegraphics[width = 1.1\hsize]{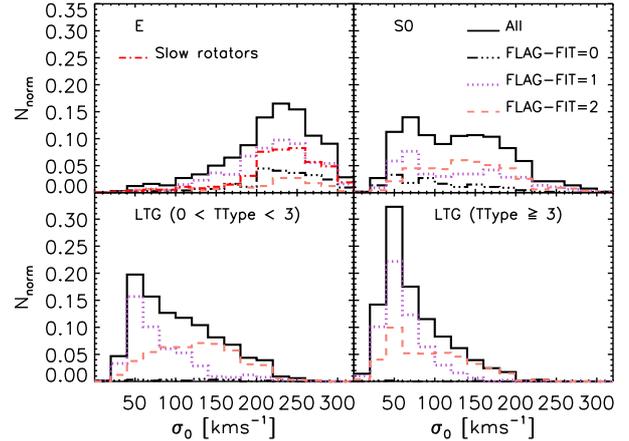}
  \vspace{-0.75cm}
  \caption{Same as previous figure, but now as a function of central velocity dispersion. LTGs with $0<$ TType $<3$ classified as having two components tend to have larger $\sigma_0$, presumably because of the bulge component.}
 \label{morphS}
\end{figure}

\begin{figure}
 \centering
  \includegraphics[width = 1.1\hsize]{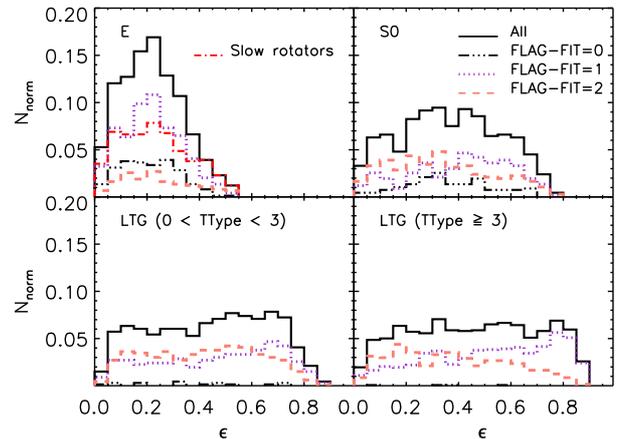}
  \vspace{-1cm}
  \caption{Same as previous figure, but now as a function of $\epsilon$.  S0s and objects with TType$>3$ tend to be rounder (peak at smaller $\epsilon$) if they are made of two components rather than one.}
 \label{morphBA12}
\end{figure}

\begin{figure*}
 \centering
 	 \includegraphics[width = 1.0\hsize]{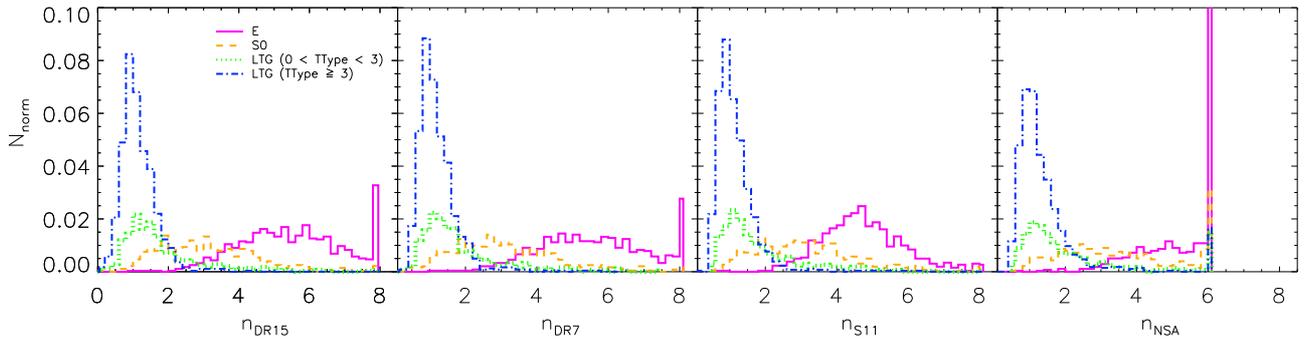}
	  \vspace{-0.5cm}
          \caption{Normalized distribution of S{\'e}rsic index $n$ for galaxies with FLAG$\textunderscore$FIT $=1$ in MPP-VAC. The distributions shown are (from left): our catalog, M15 (DR7), S11, and NSA. These histograms are divided into morphological type following MDLM-VAC: late-types TType $>3$ (blue) and $0<$ TType $<3$ (green), S0s (orange), and ellipticals (magenta). Our DR15 and DR7 analysis limits $n \le 8$, whereas the S11 analysis allows $0.5 \le n \le 8$, and NSA does not allow $n>6$. 
          }
 \label{morphn1}
\end{figure*}

\begin{figure*}
 \centering
 	 \includegraphics[width = 1.0\hsize]{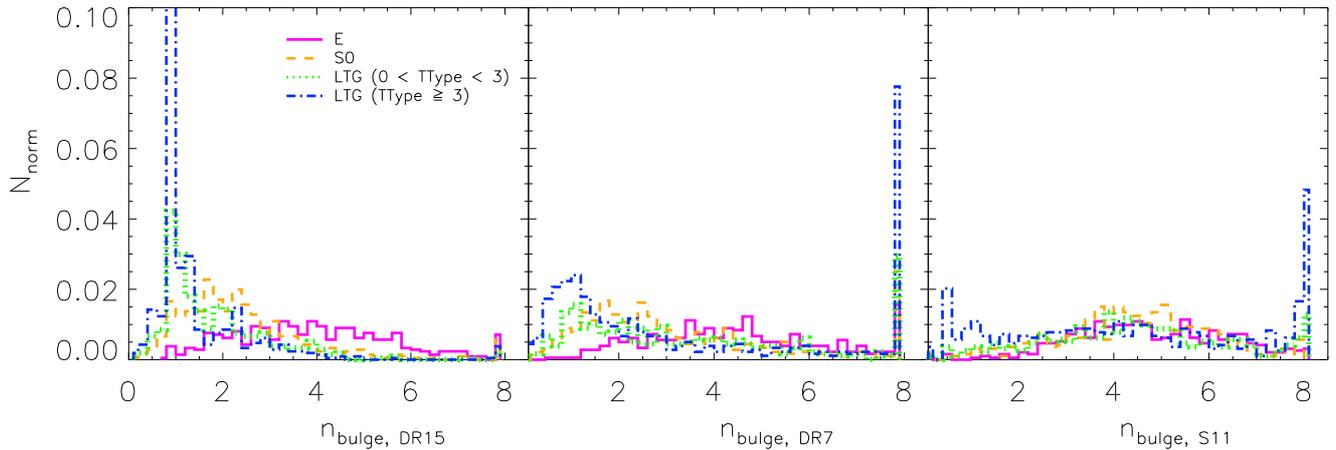}
	  \vspace{-0.5cm}
          \caption{Normalized distribution of $n$-bulge from our DR15 analysis (left), DR7 (middle), and S11 (right), for galaxies with FLAG$\textunderscore$FIT $=2$ in MPP-VAC. The histograms are divided into morphological types as in the previous figure. The spike at $n_{\rm bulge}=8$ in the middle panel (DR7) is due to late-types (not S0s); these have $n_{\rm bulge}\sim 1$ in the left-hand panel (DR15), even though the morphological classification was not used to motivate the change.}
 \label{morphn2}
\end{figure*}

\begin{figure*}
 \centering
 	 \includegraphics[width = 1.0\hsize]{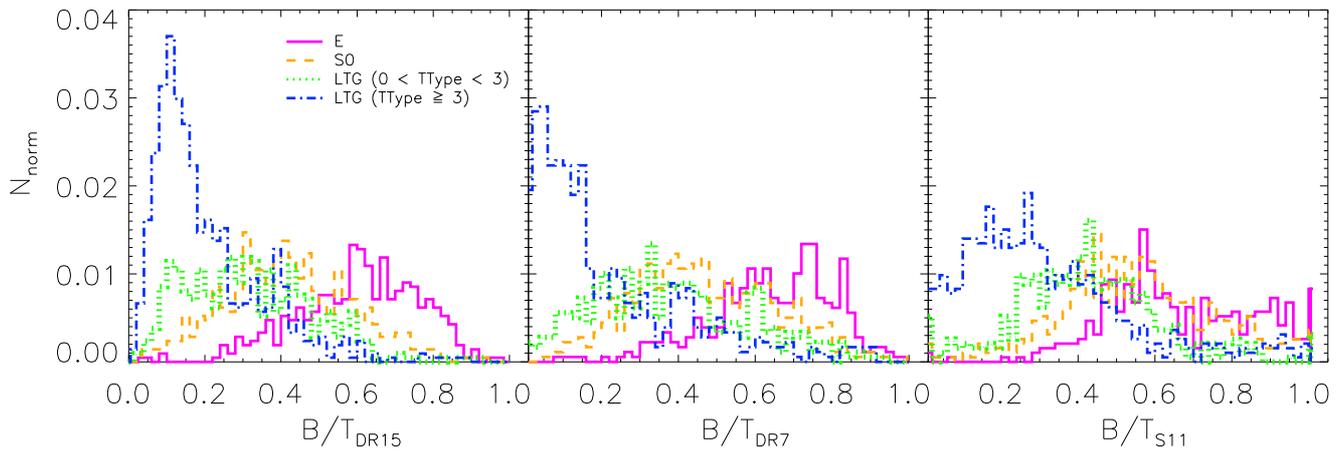}
	  \vspace{-0.5cm}
          \caption{Same as previous figure, but now for distribution of the bulge/total light ratio (B/T). The distributions from our measurements (DR15, left panel) show a clearer separation between S0s (orange) and Es (magenta) compared to S11.
          }
 \label{morphbt}
\end{figure*}

\begin{figure*}
 \centering
  \includegraphics[width = 0.7\hsize]{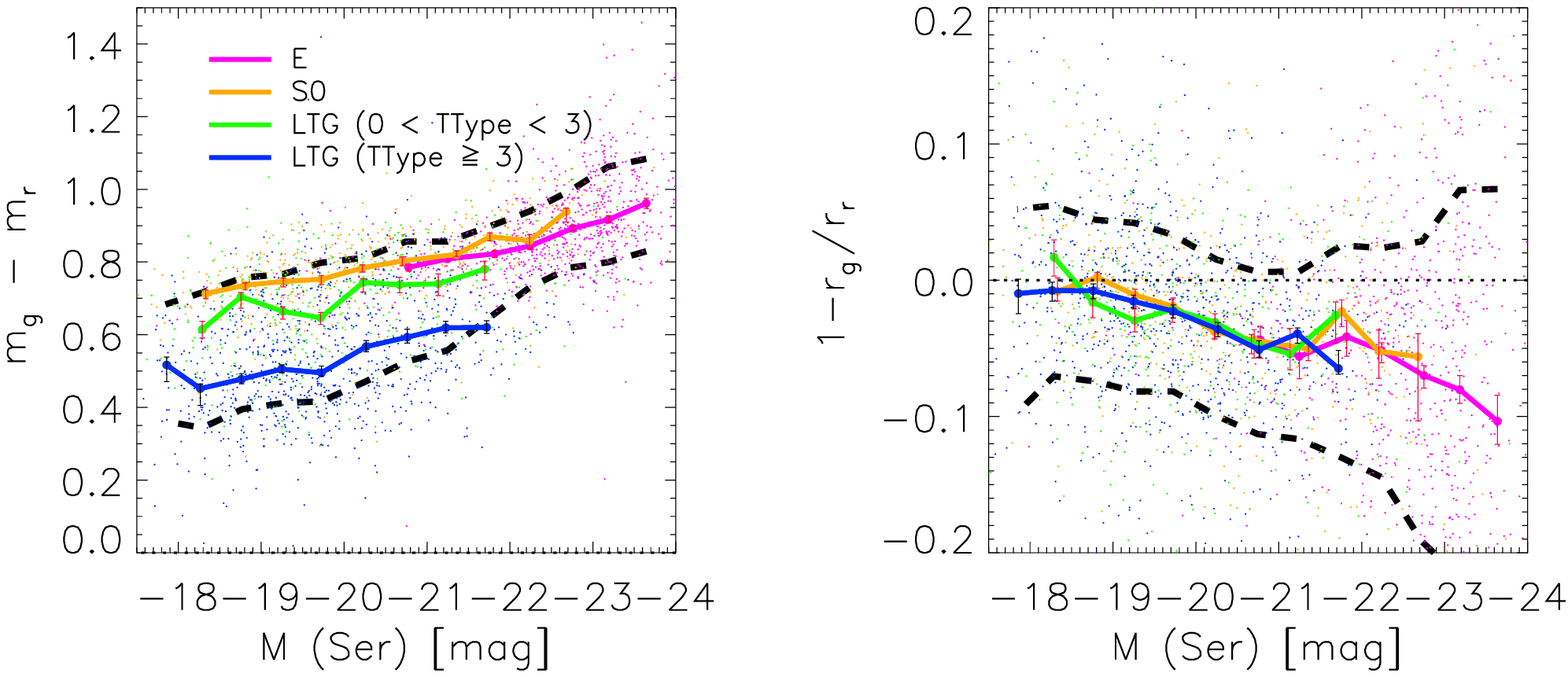}
  \includegraphics[width = 0.7\hsize]{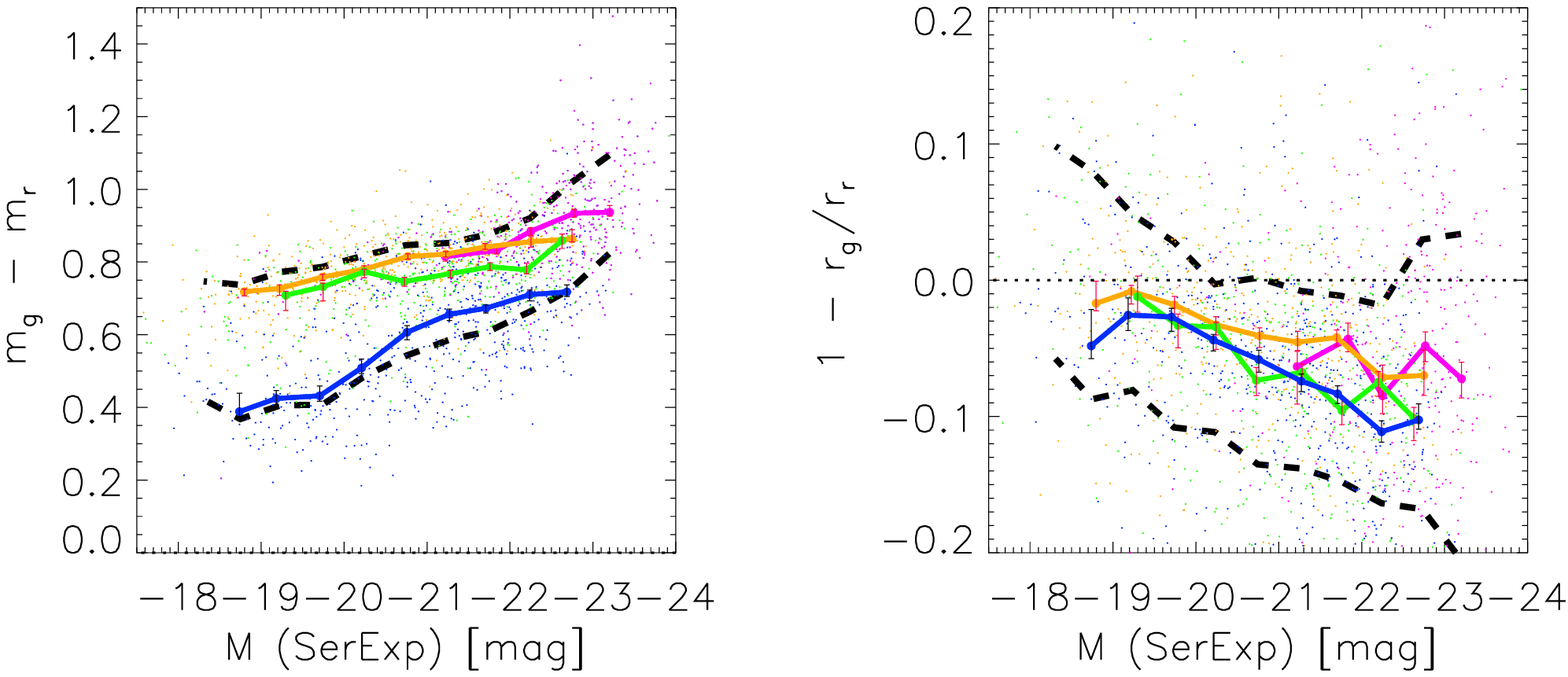}
  \vspace{0cm}
  \caption{Ratio of the total g- and r-band light (left) and size (right) as a function of morphology (blue and green represent LTGs, TType $>3$ and $0<$ TType $<3$, while orange and magenta show S0s and Es). Dashed black lines show the region which encloses 68\% of the galaxies at fixed absolute magnitude. The top panels show comparison for the Ser fit while the bottoms for the SerExp fit. The panels on the left are not color-magnitude relations in the conventional sense, as they use the total light, rather than the light within the same aperture in both bands. Note that a single-S{\'e}rsic component fit is prefered for LTG galaxies with M (Ser) $\ge -20.5$ while higher luminosity LTGs prefer a Ser-Exp fit (see also Figure~\ref{morphM} and Table~\ref{tabfit}).}
 \label{grmagsize}
\end{figure*}

\subsection{Morphology and FLAG$\textunderscore$FIT}

Figure~\ref{morphM} shows the result of dividing each morphological type into the subsets which are made of one (dotted) or two (dashed) components (i.e., FLAG$\textunderscore$FIT $=1$ or $2$) or for which both fits are equally acceptable (dot-dot-dashed; FLAG$\textunderscore$FIT = 0). (Dot-dashed histogram in top left panel shows the Es that are `slow rotators' as we discuss in Section~\ref{sec:spin}.) While the bottom part of Table~\ref{tabfit} gives the different fractions, the Figure shows quite nicely that Es with FLAG$\textunderscore$FIT = 0 tend to have high luminosities while S0s show an opposite trend (the number of LTGs with FLAG$\textunderscore$FIT = 0 is negligible).  In addition, the distribution of the absolute magnitude for Es is quite different from that in the other three panels:  Es tend to be luminous single component systems, while for S0s and LTGs, the single component systems tend to be fainter.  While the overall distribution (solid) in the other three panels is similar, the division between one- (dotted) and two- (dashed) component systems differs:
single-component LTGs with TType$>3$ are much fainter than those with two-components (bottom right); 
this difference is less apparent for $0<$TType$<3$ (bottom left); 
two-component S0s tend to be only slightly more luminous (top right).  Thus, S0s appear to be transition objects, consistent with recent work suggesting that S0s are fading spirals \citep{Rizzo2018}. 

Figure~\ref{morphS} shows that similar distributions are also seen when plotted as a function of central velocity dispersion $\sigma_0$. LTGs classified as having two components tend to have larger $\sigma_0$ -- presumably because of the bulge component. Finally, Figure~\ref{morphBA12} shows that single-component S0s and LTGs tend to have larger $\epsilon$; this is more evident for TType$>3$ consistent with them being thin disks.  (Although not the main focus of this discussion, the distribution of $\epsilon$ for Es is interesting.  We show this distribution for `slow rotators' in the top left panel of Figure~\ref{morphBA12}; it is similar for `fast rotators', except at $\epsilon>0.4$ where all Es are `fast' by definition.  Therefore, the fast rotators having $\epsilon>0.5$ in \cite{Weijmans2014} must be objects we classify as S0s.)

Overall, we believe the correspondence between FLAG$\textunderscore$FIT and morphology is remarkable, given that PyMorph played no role in the morphological classification. This is why we believe FLAG$\textunderscore$FIT contains useful information and should be used in scientific analyses of our photometric catalog.

\subsection{Morphology, S{\'e}rsic index, and B/T}
We now consider the distribution of B/T and $n$ as a function of morphological type.
We begin by showing the distribution of $n$ for our single-component galaxies (FLAG$\textunderscore$FIT $=1$). Figure~\ref{morphn1} shows that PyMorph DR15, DR7, S11, and NSA all show clear trends with $n$. These are reasonably consistent with Figure~28 of \cite{Nair2010}: LTGs tend to have $n\approx$1-2, whereas Es tend to have a broad distribution which peaks around $n\sim 5$ (recall that NSA requires $n\le 6$).

Figure~\ref{morphn2} instead shows that there are rather significant differences between the distribution of our $n$ values of the bulge component (left) and those of S11 (right) for galaxies best fitted with a SerExp profile (FLAG$\textunderscore$FIT $=2$). The middle panel shows our DR7 analysis (M15). It is worth noting that the spike at $n_{\rm bulge}=8$ in the middle panel was due to late-types (not S0s); these have $n_{\rm bulge}\sim 1$ in our DR15 analysis, even though the morphological classification was not used to motivate the re-fitting and flipping.

Comparison of the left hand panels of Figures~\ref{morphn1} and~\ref{morphn2} shows that, while there are quantitative differences, the dependence of $n_{\rm bulge}$ distribution on morphology is similar to that of $n$ on morphology for our single-component galaxies: Es (magenta) have a broad distribution centered on $n_{\rm bulge}=4$, S0s (orange) are narrower and peaked around $n_{\rm bulge}=2$, whereas LTGs (green and blue) are quite well-peaked around $n_{\rm bulge}=1$. This is impressive given that none of the PyMorph parameters played a role in the MDLM-VAC classifications. In contrast, S11 find that the distribution of $n_{\rm bulge}$ is approximately independent of morphological type. 

Figure~\ref{morphbt} shows that these differences also appear in B/T. Late-type galaxies with TType $>3$ (blue) tend to have smaller B/T values; ours tend to be peaked around B/T $\sim 0.1$ whereas S11 shows a much broader distribution. In addition, S11 find that S0s (orange) and Es (magenta) have almost the same B/T distributions, whereas our Es are clearly offset to larger B/T compared to S0s. Finally, note that LTGs with $0<$ TType $<3$ (green) are more like S0s than like TType $>3$. These trends are found despite the fact that the fitted B/T values played {\em no} role in the MDLM-VAC classifications.

To summarize, while there is general agreement that smaller B/T tends to imply a lower $n$, and this is a function of morphological type, our analysis returns a much stronger dependence of B/T and $n$-bulge on morphological type than previous work. The correspondence between photometric parameters and morphological classifications in Figures~\ref{morphMS} -- \ref{morphBA12} gives us confidence in our results.  

\subsection{Morphology and PyMorph fits in other bandpasses}\label{difbands}
Although we have mainly shown results in the $r$-band, MPP-VAC also provides PyMorph photometric parameters in the $g$- and $i$-bands. Note that the analysis in one band is independent of that in another. In contrast, for NSA and S11, $n$ is fit in $r$- and then forced to be the same in all other bands. Figure~\ref{grmagsize} shows the ratio of the total $g$- and $r$-band light (left) and size (right) as a function of morphology (blue, green, orange, and magenta represent spirals with TType $>3$ and $0<$ TType $<3$, S0s, and Es) for the objects we flag as being single-components (top) and two-components (bottom). Recall that faint LTGs and Es tend to be single-S{\'e}rsic, whereas for brighter LTGs and S0s the SerExp fit is preferred.

\begin{figure}
 \centering
  \includegraphics[width = 0.9\hsize]{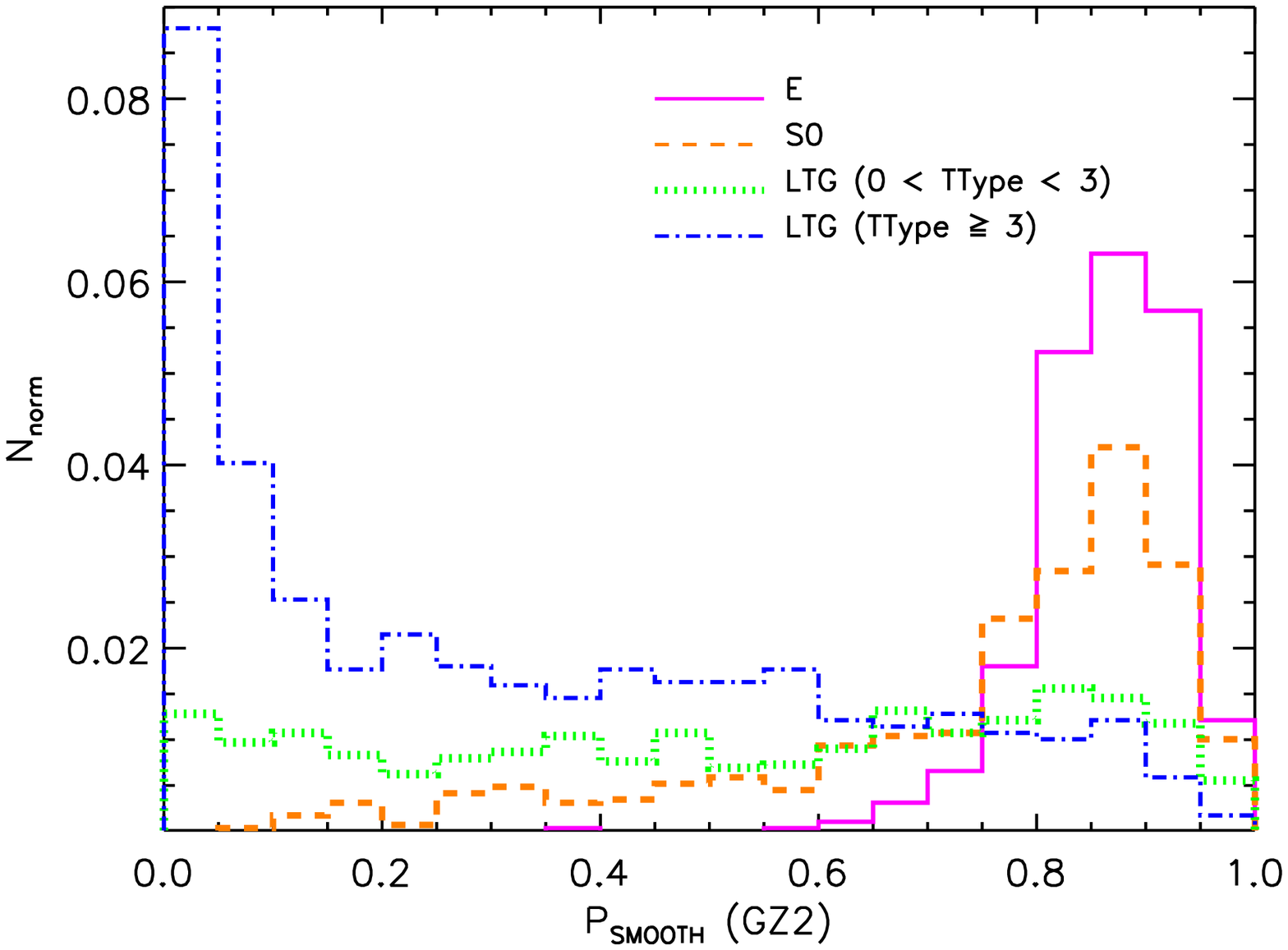}
  \includegraphics[width = 0.9\hsize]{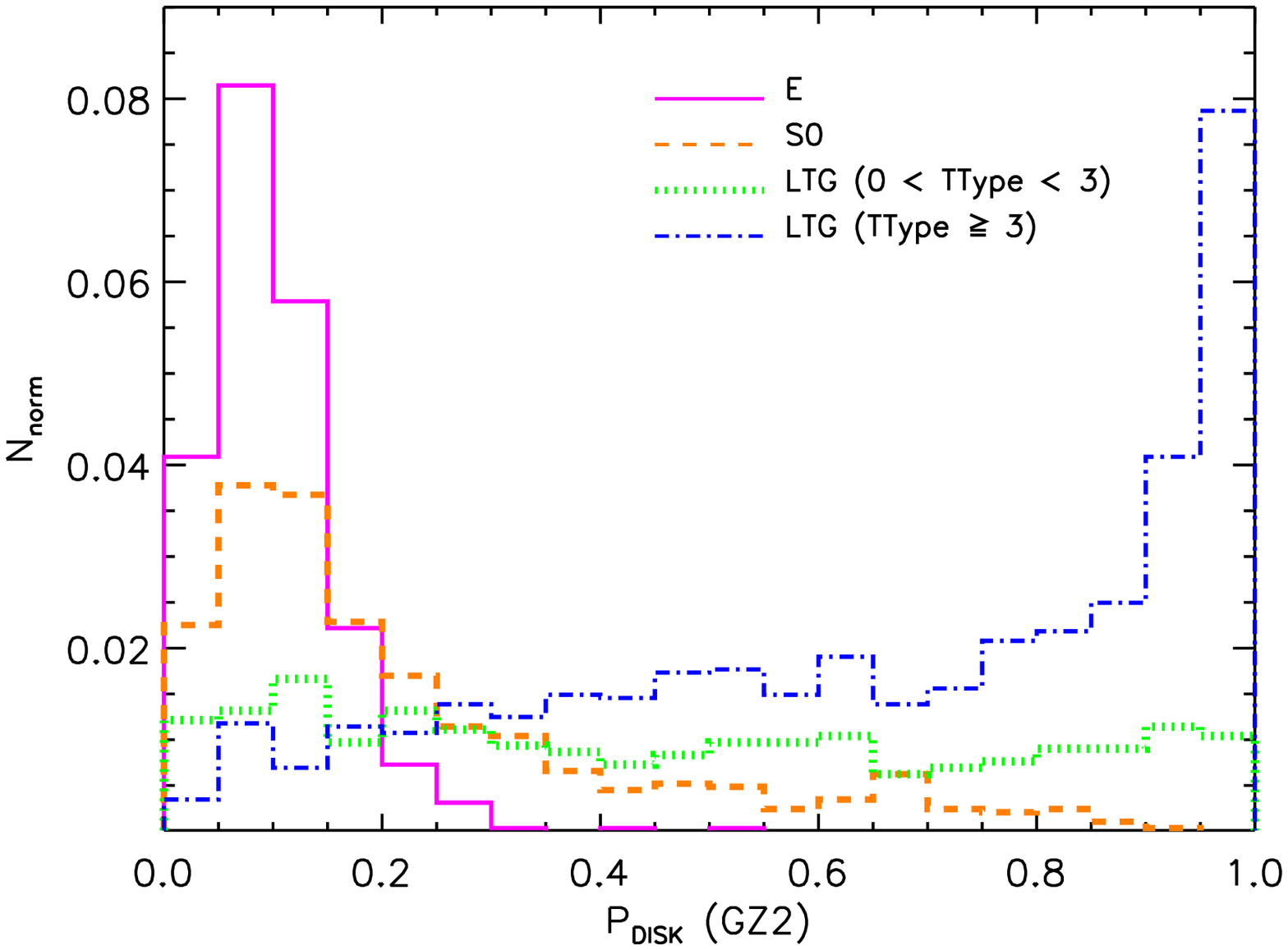}
  \vspace{-0.2cm}
  \caption{Top: Distribution of the Galaxy Zoo 2 (GZ2) probability P$_{\rm Smooth}$ for objects classified as E, S0, or LTG by our Deep Learning algorithm.  Whereas objects with P$_{\rm Smooth}<0.6$ are not contaminated by Es, a substantial fraction of objects with P$_{\rm Smooth}>0.6$ are not Es. Bottom:  Same as top, but now for the GZ2 probability P$_{\rm Disk}$.}
 \label{gz}
\end{figure}

The panels on the left are not color-magnitude relations in the conventional sense, as they use the total light, rather than the light within the same aperture in both bands. But they do show that the colors of LTGs with $0<$ TType $<3$ are more like S0s than spirals with TType $>3$. The panels on the right show that r$_{g}$ $>$ r$_{r}$ as previously observed (e.g. \citealt{B2003, R2010}), with the difference becoming larger at high luminosities, independently of morphological type. Although we do not show it here, a weak trend is also observed for the S{\'e}rsic index $n$, with high luminosity LTG and S0 galaxies having slightly larger $n$ in $r$- compared to $g$-band. No significant differences are observed for the other parameters (e.g. B/T).

\begin{figure}
 \centering
  \includegraphics[width = 0.9\hsize]{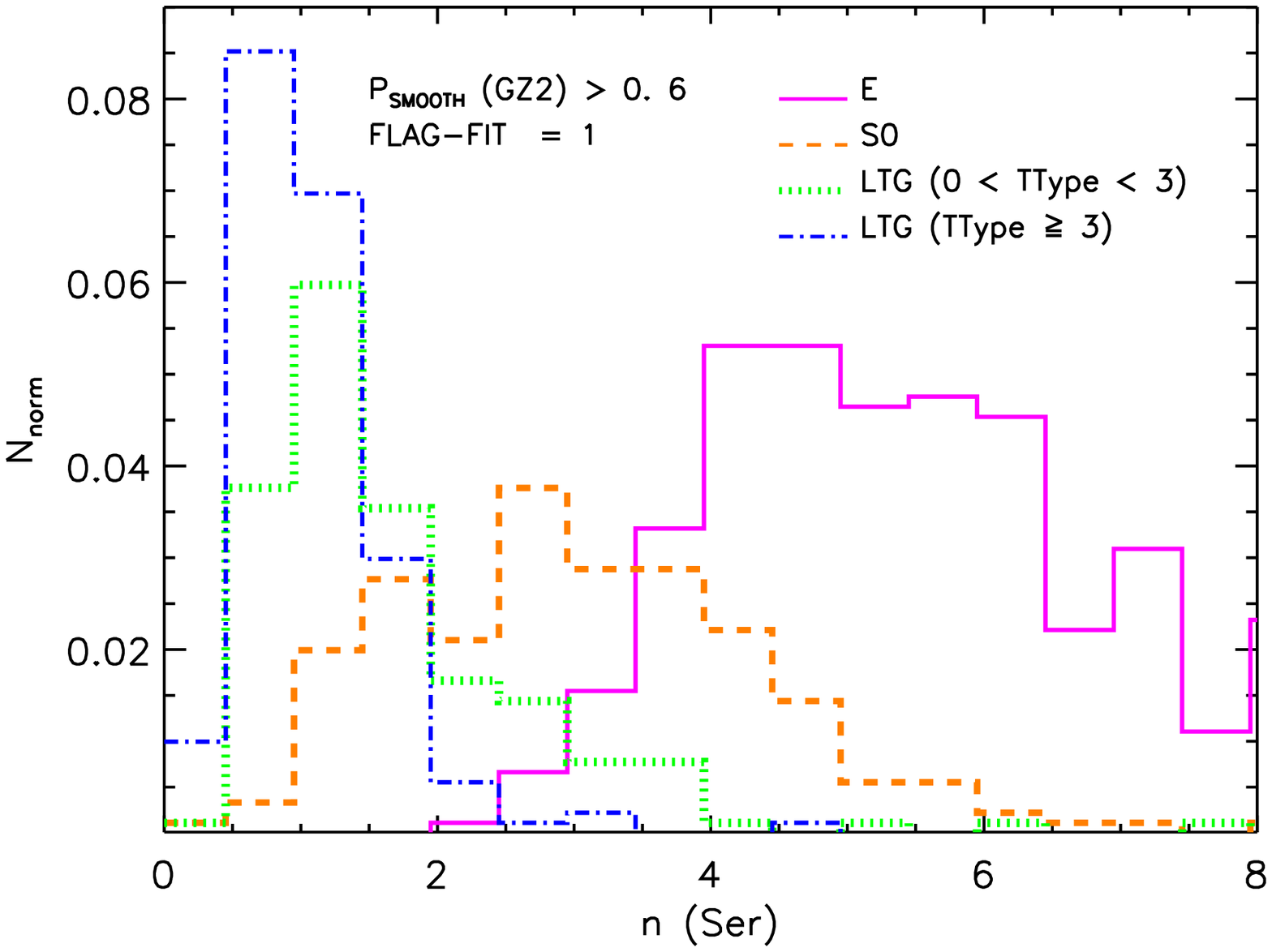}
  \includegraphics[width = 0.9\hsize]{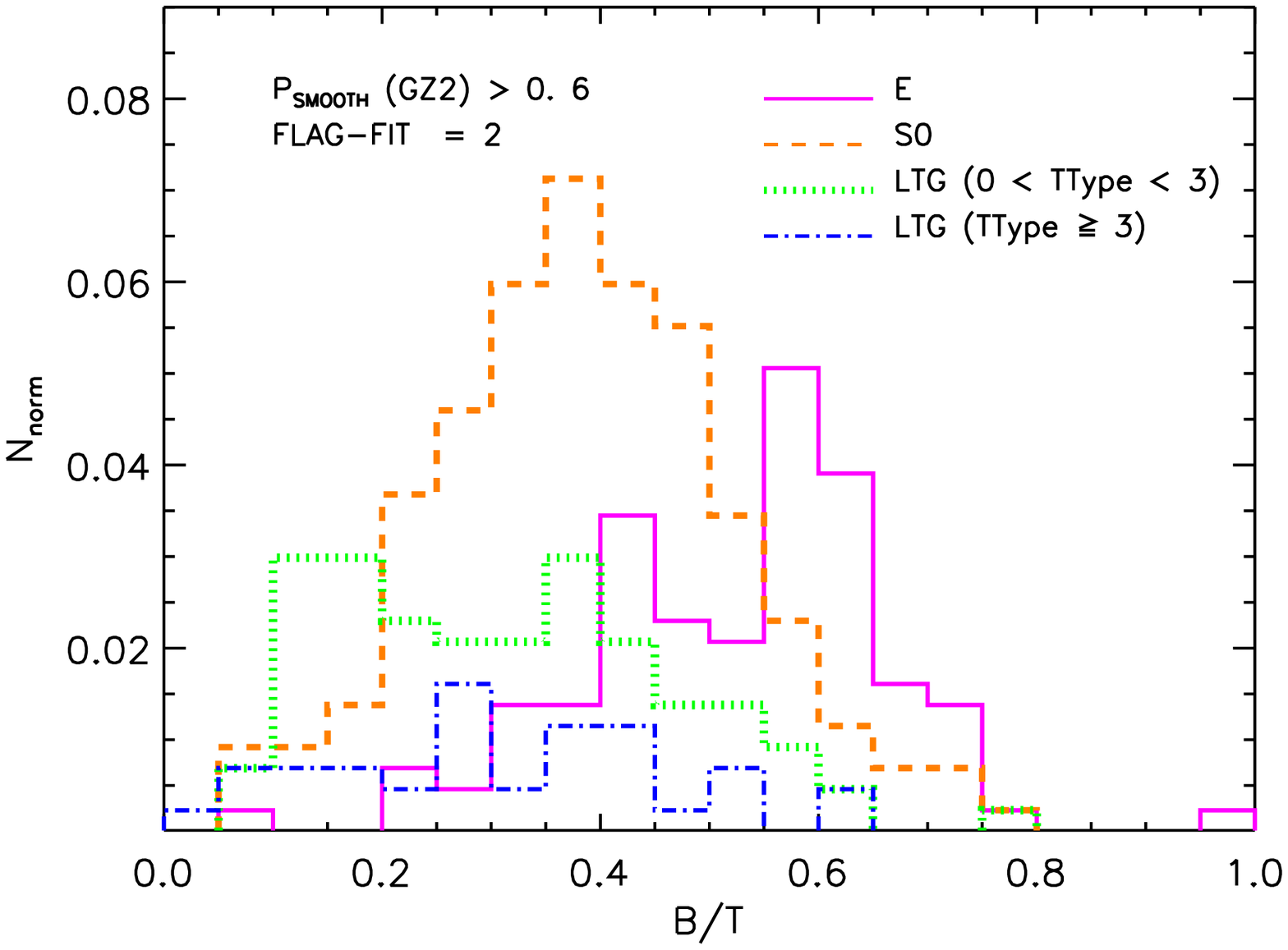}
  \vspace{-0.2cm}
  \caption{Top: Distribution of Sersic $n$ for Es, S0s and LTGs with GZ2 probability P$_{\rm Smooth}>0.6$ and FLAG$\textunderscore$FIT $=1$.  Objects with small $n$ tend to be LTGs. Bottom: Distribution of B/T for Es, S0s and LTGs having P$_{\rm Smooth}>0.6$ and FLAG$\textunderscore$FIT $=2$.  Objects with small B/T tend to be LTGs.  }
 \label{gzSmooth}
\end{figure}

\begin{figure}
 \centering
  \includegraphics[width = 0.9\hsize]{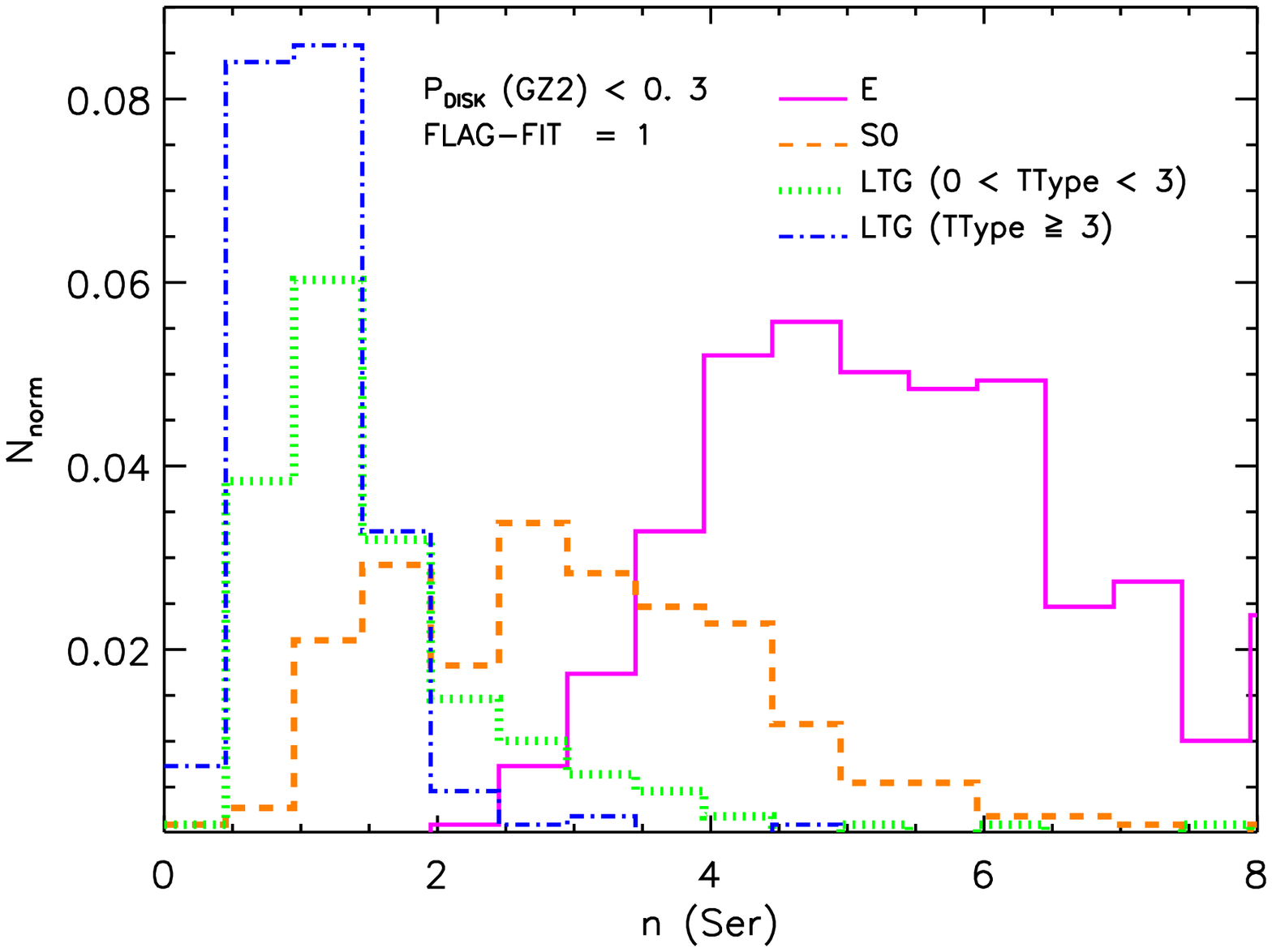}
  \includegraphics[width = 0.9\hsize]{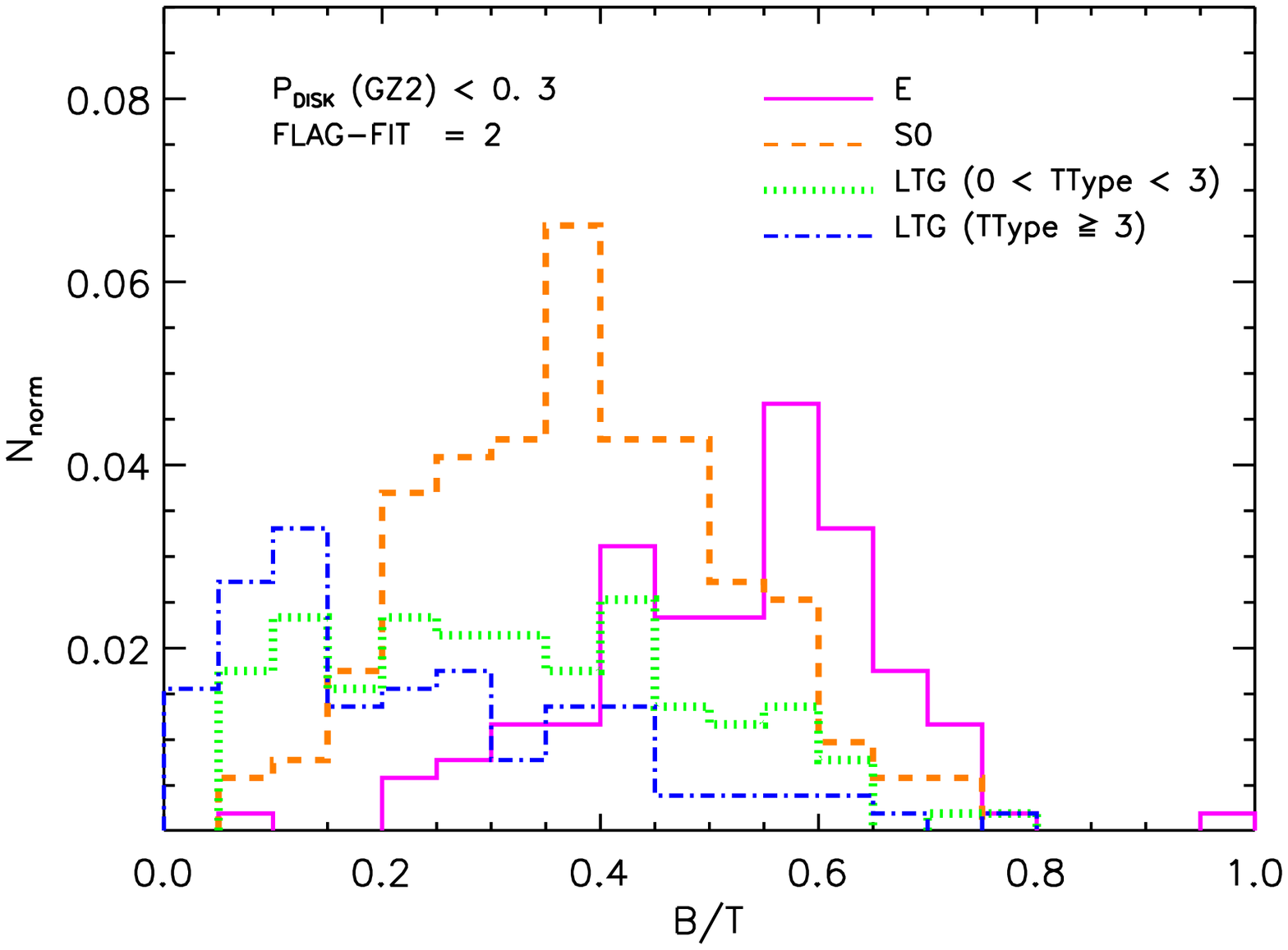}
  \vspace{-0.2cm}
  \caption{Same as previous figure, but for objects having GZ2 probability P$_{\rm Disk}<0.3$.}
 \label{gzDisk}
\end{figure}

\begin{figure*}
\centering
\includegraphics[width = 1.\hsize]{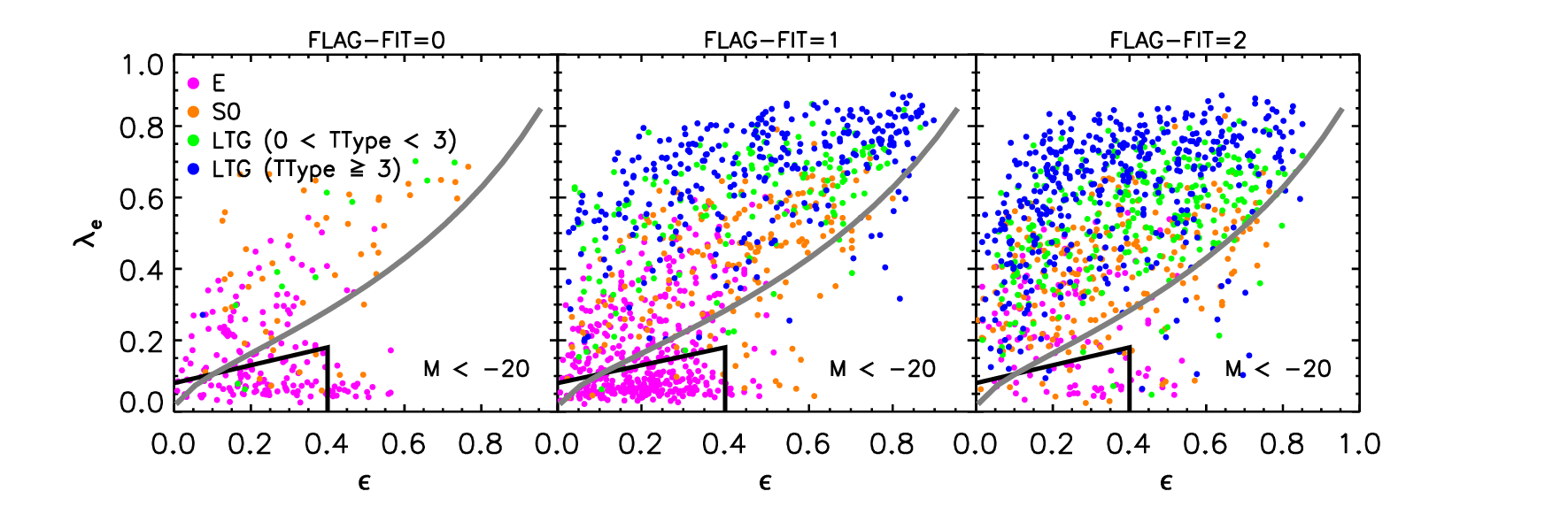}
\includegraphics[width = 1.\hsize]{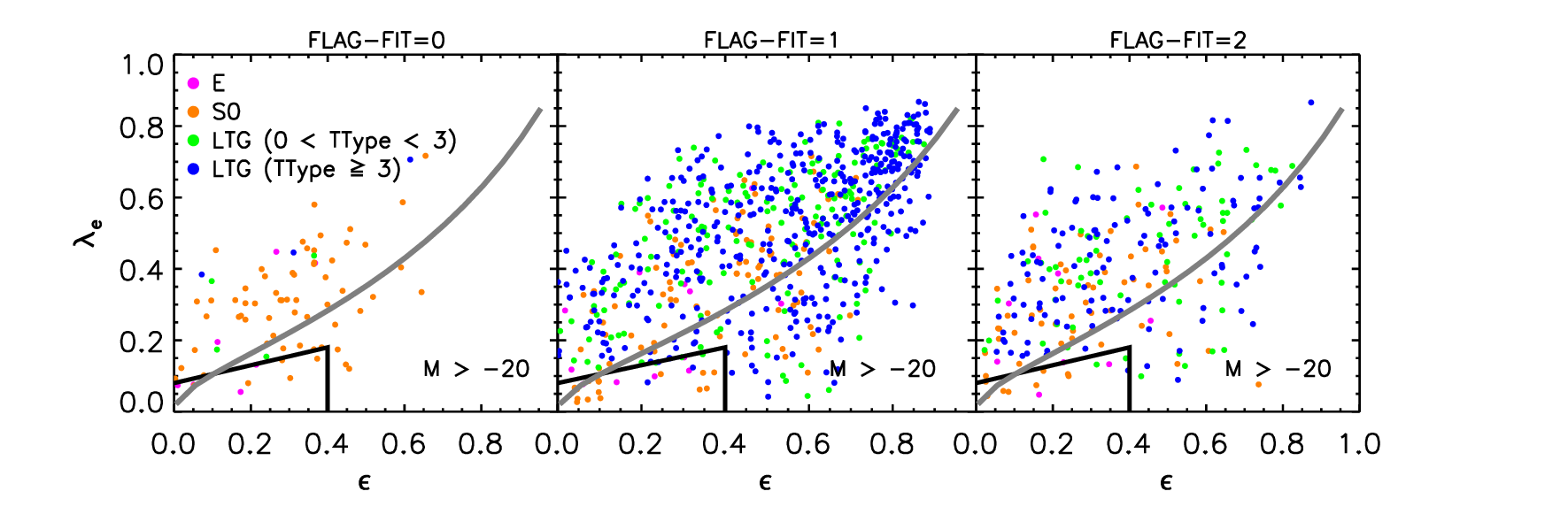}
\vspace{-0.5cm}
 \caption{Correlation between the spin parameter, $\lambda_e$, and ellipticity, $\epsilon = 1-b/a$, divided by morphological type for galaxies brighter (top) and fainter (bottom) than $M=-20$. Galaxies that are likely to be two-component systems are shown on the right, single-components in the middle, and those that are equally likely to be either in the left. The estimates of the ellipticity $\epsilon$ and absolute magnitude $M$ are from the single component S{\'e}rsic fit for galaxies with FLAG$\textunderscore$FIT $=1$, while estimates from the SerExp fit are used for galaxies with FLAG$\textunderscore$FIT $=0$ or $=2$. In each panel, magenta, orange, green and blue symbols show Es, S0s, and LTGs with TType smaller and bigger than 3. The grey curve, same in each panel, shows the result of inserting equation~(14) of C16 with $\alpha=0.15$, $\delta=0.7\epsilon_{\rm intr}$ and $i=90^\circ$ (so $\epsilon=\epsilon_{\rm intr}$) in equation~(18) of C16. The small box in the lower left corner of each panel shows the region associated with `slow rotators'; it is mainly populated by single-component Es ($\sim 45\%$ of Es are within the small box).}
\label{leBrightFaint}
\end{figure*}

\subsection{Comparison with Galaxy Zoo 2 morphologies}\label{sec:gz}
Before we move on to study correlations of spectroscopic quantities with morphology, it is interesting to contrast our MDLM Deep Learning morphologies with those of the GZ2 provided by \cite{Willett2013}.  As a first test, we use the GZ2 probabilities P$_{\rm Smooth}$ and P$_{\rm Disk}$ which are sometimes used as proxies for ``early-type" (ETGs) and ``late-type" (LTGs) galaxies.  Figure~\ref{gz} shows the distribution of P$_{\rm Smooth}$ and P$_{\rm Disk}$ values for objects which we classify as E, S0, $0<$TType$<3$ and TType$>3$.  Notice that there are no Es with P$_{\rm Smooth}<0.6$ or P$_{\rm Disk}>0.3$, so a `quasi-LTG' sample selected to have small P$_{\rm Smooth}$ or large P$_{\rm Disk}$ will not be contaminated by Es.  On the other hand, a `quasi-ETG' sample, selected to have P$_{\rm Smooth}>0.6$ or P$_{\rm Disk}<0.3$, will be strongly contaminated ($\sim 40$\%) by objects we classify as LGTs (TType$>0$).

To see if this reflects problems with the MDLM classification, we took the objects having P$_{\rm Smooth}>0.6$ and FLAG$\textunderscore$FIT $=1$ and plotted the distribution of $n$ for the subset classified as E, S0, or LTG.  For objects with P$_{\rm Smooth}>0.6$ and FLAG$\textunderscore$FIT $=2$ we did the same for B/T instead of $n$.  Figure~\ref{gzSmooth} shows the results.  There is clearly a large number of objects with $n<2$, which we classify as LTGs (TType$>0$).  Similarly, the objects which MDLM classifies as LTGs tend to have smaller B/T values than Es.  We believe this is reasonable.  Figure~\ref{gzDisk} shows a similar analysis of objects with P$_{\rm Disk}<0.3$ and FLAG$\textunderscore$FIT $=1$ or 2:  Once again, the objects classified as LTGs by MDLM have small $n$ and small B/T.  A visual inspection of these objects shows that, even though they have P$_{\rm Smooth}>0.6$ or P$_{\rm Disk}<0.3$, they really are LTGs.  Since neither $n$ nor B/T played a role in determining TType, P$_{\rm Smooth}$, or P$_{\rm Disk}$, we conclude that selecting Es based on our MDLM TType classifications is much more robust than selecting on GZ2 P$_{\rm Smooth}$ or P$_{\rm Disk}$; conclusions about Es that are based on P$_{\rm Smooth}$ or P$_{\rm Disk}$ should be treated with caution.

\section{Spectroscopy, photometry, and morphology: Stellar angular momentum}\label{sec:spin}
In the previous section, we described how PyMorph photometry can be combined with morphology. Here, we combine photometry, morphology, and spectroscopy to study the stellar angular momentum in MaNGA galaxies.

The next series of figures show how the correlation between the spin parameter $\lambda_e$ defined by \cite{Emsellem2007}, and ellipticity, $\epsilon \equiv 1-b/a$, depends on luminosity, velocity dispersion, and morphological type. To do so, we measure
\begin{equation}
  \lambda_e \equiv \frac{\sum_i^N R_i F_i |V_i|}{\sum_i^N R_i F_i \sqrt{V_i^2 + \sigma_i^2}},
  \label{lambdae}
\end{equation}
where $R$, $F$, $V$ and $\sigma$ denote the circularized radius, flux, rotational velocity and velocity dispersion of the $i$-th spaxel.  The sum is over all $N$ spaxels within elliptical isophotes, out to the half-light radius (returned by S{\'e}rsic or SerExp fits, depending on FLAG$\textunderscore$FIT), which we then PSF-correct following \cite[][hereafter G18]{Graham2018}. 
For the discussion which follows, it is useful to also define 
\begin{equation}
 \langle\sigma^2\rangle_e \equiv \frac{\sum_i^N F_i \sigma_i^2}{\sum_i^N F_i}.
\end{equation}
and
\begin{equation}
 \langle v_{\rm rms}^2\rangle_e \equiv \frac{\sum_i^N F_i (V_i^2 + \sigma_i^2)}{\sum_i^N F_i} = \langle V^2\rangle_e + \langle\sigma^2\rangle_e.
\end{equation}
Finally, we use $\sigma_e$ and $V_e$ to denote the value of the dispersion and rotational speed at $R_e$, respectively ($\sigma_e^2$ is almost always smaller than $\langle\sigma^2\rangle_e$, the light-weighted value of the dispersion within $R_e$).

\begin{figure*}
 \centering
  \includegraphics[width = 1.0\hsize]{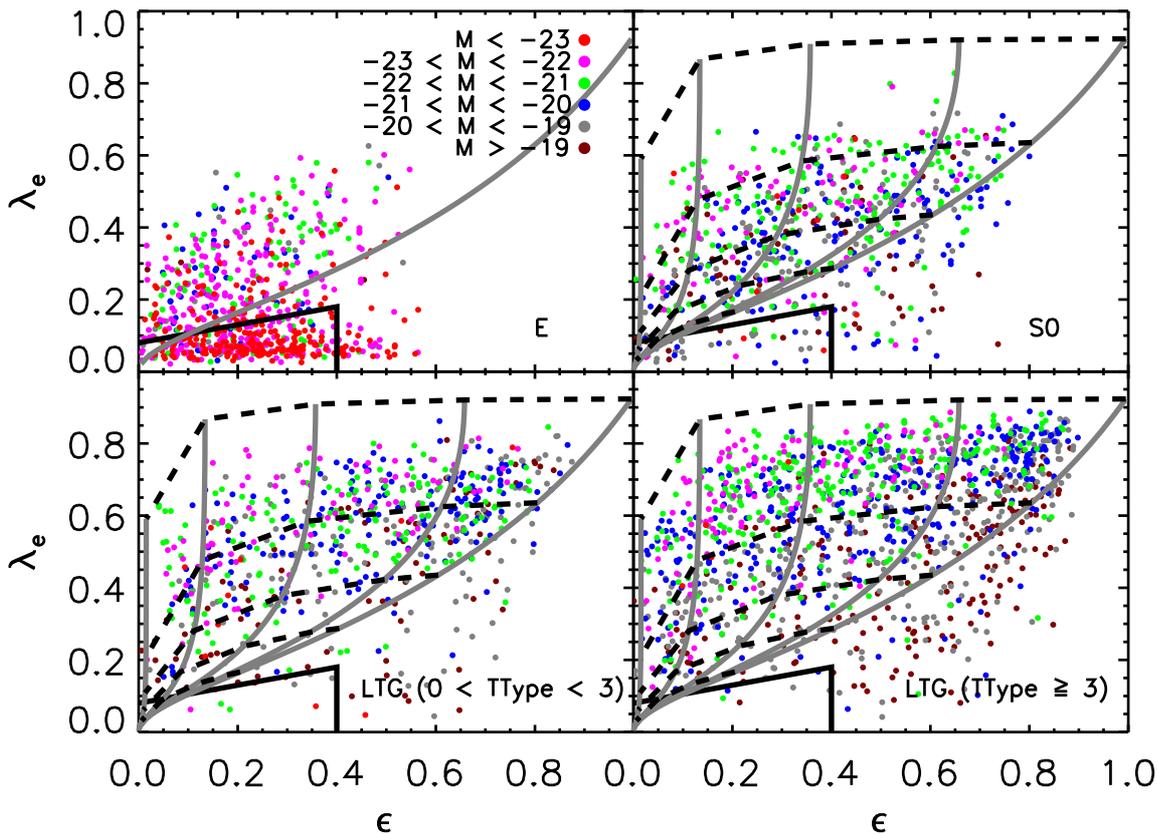}
 \vspace{-1.5cm}
  \caption{Correlation between $\lambda_e$ and $\epsilon$, as a function of morphological type and total absolute magnitude. The estimates of the ellipticity $\epsilon$ and absolute magnitude $M$ are from the single component S{\'e}rsic fit for galaxies with FLAG$\textunderscore$FIT $1$, while estimates from the SerExp fit are used for galaxies with FLAG$\textunderscore$FIT $=0$ or $=2$. Smooth grey curves, same in each panel, show the result of inserting equation~(14) of C16 with $\alpha=0.15$, $\delta=0.7\epsilon_{\rm intr}$ and $i=90^\circ - j\, 20^\circ$ with $j=[0,4]$ in equation~(18) of C16.  Dashed curves show lines of fixed $\epsilon_{\rm intr} = 1 - 0.2j$ with $j=[0,4]$.  Later morphological types tend to have larger $\lambda_e$.  Whereas fainter LTGs have lower $\lambda_e$ (bottom right), the trend with luminosity is opposite for Es (upper left).  Indeed, luminous Es dominate at small $\lambda_e$ and $\epsilon$, whereas luminous LTGs dominate at large $\lambda_e$. }
 \label{leL}
\end{figure*}

\begin{figure*}
 \centering
  \includegraphics[width = 0.9\hsize]{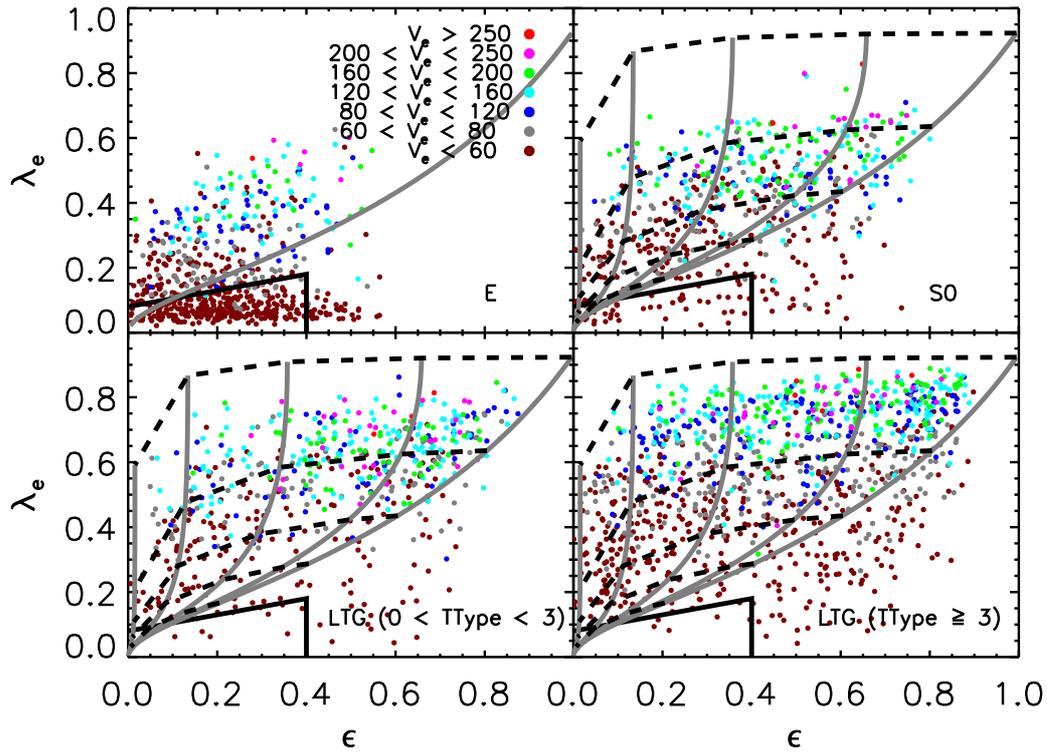}
  \vspace{-1.5cm}
  \caption{Same as previous figure, but now as a function of rotational velocity $V_e$ on the scale $R_e$. Galaxies with the largest $V_e$ have larger $\lambda_e$ as TType increases. }
 \label{leVrot}
\end{figure*}

\begin{figure*}
 \centering
  \includegraphics[width = 0.9\hsize]{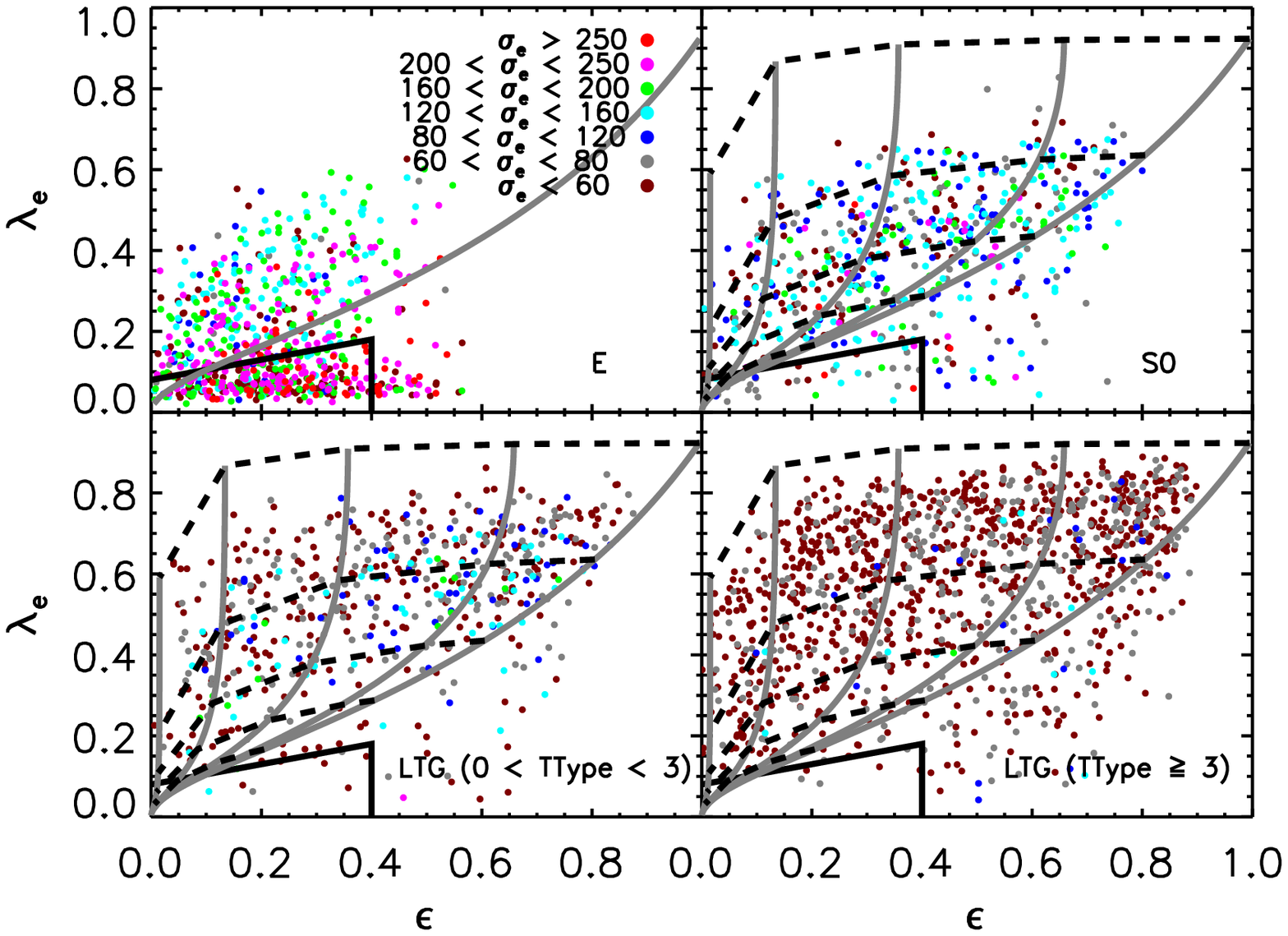}
  \vspace{-1.5cm}
  \caption{Same as previous figure, but now as a function of $\sigma_e$, the velocity dispersion at (not within) the scale $R_e$. The typical $\sigma_e$ decreases as TType increases.}
 \label{leSate}
\end{figure*}

We use the estimates of stellar rotational velocity and velocity dispersion from the MaNGA 3D kinematics maps (Westfall et al. 2019, in preparation).
In practice, we only include in the sum spaxels having S/N$>$5 (although increasing the cut to S/N$>$8 or 10 makes no significant difference for the $\lambda_e$-related results which follow), STELLAR$\textunderscore$VEL$\textunderscore$MASK=0 and STELLAR$\textunderscore$SIGMA$\textunderscore$MASK=0. The velocity dispersion $\sigma_i$ is corrected for instrumental resolution. (To account for the difference in resolution between the MILES templates and the MaNGA data, the STELLAR$\textunderscore$SIGMACORR values must be subtracted in quadrature from STELLAR$\textunderscore$SIGMA (Westfall et al. 2018, in preparation): i.e. 
\begin{equation}
\sigma_i^2 = {\rm STELLAR\_SIGMA}^2_i - {\rm STELLAR\_SIGMACORR}^2_i,
\label{si}
\end{equation}
with median(STELLAR$\textunderscore$SIGMACORR)~$\sim 32$~kms$^{-1}$. STELLAR$\textunderscore$SIGMA $<$ STELLAR$\textunderscore$SIGMACORR for some spaxels; for these, we simply set $\sigma_i = 0$ kms$^{-1}$.  Setting $\sigma_i$ for these spaxels to be as large as $20$ kms$^{-1}$ makes little difference to the results which follow.)

Figure~\ref{leBrightFaint} shows results for galaxies that are brighter (top) and fainter (bottom) than $M = -20$. From right to left, the three panels are for galaxies classified as two-component systems (FLAG$\textunderscore$FIT $=2$), single (FLAG$\textunderscore$FIT $=1$), or either (FLAG$\textunderscore$FIT $=0$), respectively. In each panel, magenta, orange, green and blue symbols show Es, S0s, and LTGs having $0\le {\rm TType}<3$ and ${\rm TType}>3$. The grey curve, same in each panel, shows the result of inserting equation~(14) of \citet[hereafter C16]{C16} with $\alpha=0.15$, $\delta=0.7\epsilon_{\rm intr}$ and $i=90^\circ$ (so $\epsilon=\epsilon_{\rm intr}$) in equation~(18) of C16. It represents a galaxy viewed edge on with velocity anisotropy parameter $\delta$, and the curve serves mainly to guide the eye. The small box in the lower left corner of each panel shows the region associated with `slow rotators' (equation~19 of C16).

The vast majority of Es appear in the upper middle panel: luminous single-component Es account for most of the `slow rotators'. S0s and LTGs with $0<$ TType $<3$ tend to have FLAG$\textunderscore$FIT $=2$, and almost the same distribution in all the panels, with the LTGs having slightly larger $\lambda_e$. In contrast, objects with TType $>3$ tend to have the largest $\lambda_e$, if they are luminous. Faint LTGs with TType $>3$ tend to be single component systems (see also Figure~\ref{morphM} and Table~\ref{tabfit}).

In general, our results are consistent with the analysis of G18, but there are a few important differences due to improvements in the spectral resolution estimate between the SDSS-DR14 and DR15 reductions (equation~\ref{si} -- see Westfall et al. 2018 in preparation for details), and in the morphological classifications.  For example, the left hand panel of Figure~8 in G18 shows many more objects with $\lambda_e>0.8$ than we find.  The difference is most pronounced at $\epsilon<0.2$, where we have almost no objects with $\lambda_e>0.8$ (our results are in better agreement with those of \citealt{Lee2018}). Another striking difference is seen in the distribution of S0s and spirals: the top right corner of our $\lambda_e-\epsilon$ plane is dominated by spirals (this is more evident in Figure~\ref{leL}); the top right corner of G18's Figure~8 is dominated by S0s. In addition, for us, the lower right corner is dominated by lower luminosity LTGs -- the majority with FLAG$\textunderscore$FIT $=1$ (bottom middle panel).  

Figure~9 of G18 shows that the mean stellar mass of galaxies in a bin of $\lambda_e$ and $\epsilon$ is approximately proportional to $\lambda_e + \epsilon$, with larger luminosities having smaller $\lambda_e+\epsilon$. We do not see this.
To explore this further, Figure~\ref{leL} shows the distribution in the $\lambda_e-\epsilon$ plane for fixed morphological type, further subdivided by luminosity. The estimates of the ellipticity $\epsilon$ and absolute magnitude $M$ are from the single component S{\'e}rsic fit for galaxies with FLAG$\textunderscore$FIT $=0$ or $=1$, while estimates from the SerExp fit are used for galaxies with FLAG$\textunderscore$FIT $=2$.
To ease comparison between panels, the smooth grey curves, same in each panel, show the result of inserting equation~(14) of C16 with $\alpha=0.15$, $\delta=0.7\epsilon_{\rm intr}$ and $i=90^\circ - j\, 20^\circ$ with $j=[0,4]$ in equation~(18) of C16. Dashed curves (same in all but top left panel) show lines of fixed $\epsilon_{\rm intr} = 1 - 0.2j$ with $j=[0,4]$.

The first point to note is that the upper most black dashed line shows the $\epsilon_{\rm intr}=1$ limit:  there should be {\em no} galaxies with small (observed) $\epsilon$ and large $\lambda_e$, and indeed, we see none.  Second, the upper envelope of the distribution increases systematically with increasing TType (compare different panels), consistent with the expectation that later types are more rotationally supported. This clear and reasonable trend with morphology is not evident in G18.  Third, the top left panel shows that the most luminous Es are slow rotators, and fainter Es have larger $\lambda_e$.  While this is consistent with G18, the upper envelope in $\lambda_e$ for the faster rotating Es is similar to that for S0s:  in contrast, for G18, S0s can have very large $\lambda_e$.  Finally, the luminosity dependence (which is evident for Es) is absent or inverted for S0s and LTGs with $0<$ TType $<3$, and is clearly inverted for LTGs with TType $>3$.   


We have also colored objects by their rotation speed $V_e$ (Figure~\ref{leVrot}) or velocity dispersion $\sigma_e$ (Figure~\ref{leSate}). We use $\sigma_e$, the velocity dispersion on the scale $R_e$, rather than $\langle\sigma^2\rangle_e$ (which is rarely used) or $\langle v_{\rm rms}^2\rangle_e$, which was used by the SAURON and Atlas3D collaborations \citep{Cappellari2006}.  The top left panel of Figure~\ref{leVrot} shows that the Es that are slow rotators have small $V_e$; the corresponding panel in Figure~\ref{leSate} shows they also have large $\sigma_e$. I.e., $\lambda_e$ is small both because the numerator in equation~(\ref{lambdae}) is small and because the denominator is large. From S0s to LTGs, the objects with the largest rotation speeds, $V_e \sim 160$~km~s$^{-1}$, have $\lambda_e$ increasing with TType because the velocity dispersion $\sigma_e$ is decreasing. (In this context, notice that the fraction of S0s with small $\sigma_e<80$~km$s^{-1}$ is much lower than for objects having $0<$TType$<3$.  The morphological dependence is stronger than for the central $\sigma_0$ shown in Figure~\ref{morphS}, but this is not unexpected, since Figure~\ref{morphbt} shows that S0s have larger B/T.)  Finally, the objects in the bottom right corner of the TType $>3$ panel have small $V_e$ {\em and} small $\sigma_e$. Noise and resolution effects mean that $\lambda_e$ for these objects may be biased. Note, however, that they approximately overlap the objects with $60< \sigma_e < 80$~km~s$^{-1}$, which we believe are reliable.  

The tendency for $\sigma_e$ to decrease systematically as TType increases (compare typical colors in the panels of Figure~\ref{leSate}) is remarkable, as neither $V_e$ nor $\sigma_e$ played any role in the morphological classification. 

\section{Conclusions}\label{final}
We presented the contents of MPP-VAC -- the PyMorph Ser and SerExp photometric structural parameters of MaNGA galaxies in the $g$, $r$, and $i$ bands (Table~\ref{Table1}) -- and its sister catalog MDLM-VAC (Table~\ref{morphcatcontents}), which provides Deep-Learning derived morphologies for the SDSS-DR15 MaNGA sample. 

Each object in MPP-VAC has a flag, FLAG$\textunderscore$FIT, which indicates the preferred set of photometric parameters that should be used for unbiased scientific analyses. We showed that the parameters from a single-S{\'e}rsic fit are in good agreement with those in the NSA (Figure~\ref{SerMag}). However our estimates, and those of the NSA, differ more significantly from those of S11.  Discussion in the recent literature suggests our estimates are more reliable because they include a more careful treatment of the background sky level.  For two-component fits, we were only able to compare our SerExp parameters with S11 (Figure~\ref{litSerExp}) because the NSA catalog does not provide two-component fits. Our DR15 SerExp photometry returns slightly less light, smaller sizes, smaller $n$-bulge, and smaller B/T. Most of these differences are driven by the relatively large offset in $n$. 

Section~\ref{sec:gz} argued that the morphological classifications from our MDLM-VAC (Section~\ref{sec:morph}) are more accurate to those from the GZ2, especially for selecting ``early-type'' galaxies. While a ``late-type'' sample selected to have small GZ2 P$_{\rm Smooth}$ or large GZ2 P$_{\rm Disk}$ will not be contaminated by Es, an ``early-type'' sample, selected to have GZ2 P$_{\rm Smooth}>0.6$ or GZ2 P$_{\rm Disk}<0.3$, will be strongly contaminated ($\sim 40$\%) by objects we classify as LTGs (TType$>0$) (Figure~\ref{gz}--\ref{gzDisk}). In addition, we discuss how the MDLM-VAC parameters can be used to divide ``early-type'' galaxies into Es and S0s (see Figures~\ref{morphMS} -- \ref{morphBA12}). Our results suggest that S0s appear to be transition objects, consistent with recent work suggesting that S0s are fading spirals.

As a simple illustration of the analysis which MPP-VAC enables, we combined it with the MDLM-VAC (Section~\ref{sec:photmorph}). We showed that the parameters returned by our two-component fits (e.g. bulge-total light ratio, bulge S{\'e}rsic index) exhibit interesting correlations with morphological type -- correlations which are absent in previous work (e.g. S11). For example, while it is known that single-S{\'e}rsic fits to late-type and early-type galaxies are likely to return S{\'e}rsic indices of $n \le 2$ and $\ge 4$, some literature suggests that there is little correlation between the S{\'e}rsic index of the bulge component and the morphology of these galaxies (Figures~\ref{morphn2} -- \ref{morphbt}). We find a correlation, despite the fact that MPP-VAC photometry and MDLM-VAC morphological determinations were performed independently. 

As another example, Section~\ref{sec:spin} presented a simple analysis of the angular momentum of MaNGA galaxies. This combines the photometric information in MPP-VAC and the morphological classifications in MDLM-VAC with independent spatially resolved spectroscopic information provided by MaNGA IFUs. We again find strong correlations with morphology (Figure~\ref{leL}) which were not present in previous work (e.g. G18).  
We also find $\lambda_e$ (equation~\ref{lambdae}) is more strongly correlated with rotation speed at the half-light radius than it is with the velocity dispersion on this scale (Figures~\ref{leVrot} and~\ref{leSate}). There is a strong tendency for $\sigma_e$ to decrease systematically as TType increases (Figure~\ref{leSate}). In addition, for Es, $\lambda_e$ decreases as the velocity dispersion $\sigma_e$ increases. In general, galaxies with LTGs with TType $> 3$ have $\sigma_e < 80$~kms$^{-1}$.

All the observed trends discussed in this paper between stellar kinematics, photometric properties, and morphological type are impressive given that the PyMorph parameters, the MDLM-VAC classifications and the spatially resolved spectroscopic parameters are totally independent estimates.

The MPP-VAC and its sister catalog MDLM-VAC are part of SDSS-DR15 and are available online$^{3,4}$ from the SDSS IV website (DR15 release$^2$). 
We expect the parameters provided in MPP-VAC and MDLM-VAC to enable a wide variety of analyses.


\subsection*{Acknowledgements}
We are grateful to the referee for a helpful report, to M. Graham, M. Huertas-Company and A. Meert for their assistance, and to R. K. Sheth for many helpful discussions. This work was supported in part by NSF grant AST-1816330.

Funding for the Sloan Digital Sky Survey IV has been provided by the Alfred P. Sloan Foundation, the U.S. Department of Energy Office of Science, and the Participating Institutions. SDSS acknowledges support and resources from the Center for High-Performance Computing at the University of Utah. The SDSS web site is www.sdss.org.

SDSS is managed by the Astrophysical Research Consortium for the Participating Institutions of the SDSS Collaboration including the Brazilian Participation Group, the Carnegie Institution for Science, Carnegie Mellon University, the Chilean Participation Group, the French Participation Group, Harvard-Smithsonian Center for Astrophysics, Instituto de Astrof{\'i}sica de Canarias, The Johns Hopkins University, Kavli Institute for the Physics and Mathematics of the Universe (IPMU) / University of Tokyo, Lawrence Berkeley National Laboratory, Leibniz Institut f{\"u}r Astrophysik Potsdam (AIP), Max-Planck-Institut f{\"u}r Astronomie (MPIA Heidelberg), Max-Planck-Institut f{\"u}r Astrophysik (MPA Garching), Max-Planck-Institut f{\"u}r Extraterrestrische Physik (MPE), National Astronomical Observatories of China, New Mexico State University, New York University, University of Notre Dame, Observat{\'o}rio Nacional / MCTI, The Ohio State University, Pennsylvania State University, Shanghai Astronomical Observatory, United Kingdom Participation Group, Universidad Nacional Aut{\'o}noma de M{\'e}xico, University of Arizona, University of Colorado Boulder, University of Oxford, University of Portsmouth, University of Utah, University of Virginia, University of Washington, University of Wisconsin, Vanderbilt University, and Yale University.

\bibliographystyle{mnras}
\bibliography{biblio} 


\end{document}